\documentclass[aps,prd,twocolumn,nofootinbib]{revtex4-2}
\usepackage{graphicx}
\usepackage[utf8]{inputenc}
\usepackage{placeins}
\usepackage{amssymb, amsmath}
\usepackage{tabularx}
\usepackage[normalem]{ulem}
\usepackage{pbox}

\newcommand{\Mthresh}{$M_{\rm thresh}$}
\newcommand{\Msub}{$M_{\rm sub}$}
\newcommand{\Msup}{$M_{\rm sup}$}
\newcommand{\Mtov}{$M_{\rm max}^{\rm TOV}$}

\newcommand{\Mtot}{$M_{\rm tot}$}
\newcommand{\amin}{$\alpha_{\rm min}$}

\newcommand{\Ms}{M_{\odot}}

\newcommand{\rns}{\rho_{\rm sat}}

\newcommand{\Rsmall}{$R_{1.4}\approx10.6$~km\ }
\newcommand{\Rbig}{$R_{1.4}\approx14$~km\ }

\usepackage[dvipsnames, usenames]{xcolor}

\newcommand{\coderefs}{\cite{Duez:2005sg,Paschalidis:2010dh,Etienne:2011re,Etienne:2015cea}}

\begin{document}

\title{Impact of thermal effects on prompt-collapse binary neutron star mergers}
\author{Carolyn A. Raithel,$^{1}$ Vasileios Paschalidis,$^{2,3}$}
\affiliation{$^1$Department of Physics and Astronomy, Swarthmore College, Swarthmore PA 19081, USA}
\affiliation{$^2$Department of Astronomy and Steward Observatory, University of Arizona, Tucson AZ 85721, USA}
\affiliation{$^3$Department of Physics, University of Arizona, Tucson AZ 85721, USA}

\begin{abstract}
The fate of the remnant following the merger of two neutron stars
initially on quasicircular orbits depends primarily on the mass of the
initial neutron stars, the mass ratio, and the still-uncertain
dense-matter equation of state (EoS).  Previous works studying the
threshold mass for prompt collapse to a black hole have primarily
focused on the uncertainties in the zero-temperature EoS, which are
parametrized by a macroscopic quantity such as the characteristic
neutron star radius. However, prompt collapse can take place either
with or without a core bounce during the merger. In the
bounce-collapse scenario, shocks can produce additional thermal
support, potentially altering the threshold for collapse. In this
work, we investigate the impact of the uncertainties in the
finite-temperature part of the nuclear EoS on the threshold mass for prompt
collapse in equal mass mergers. Using two cold EoSs, combined with four parametrizations of
the finite-temperature part of the EoS, we find that the threshold
mass is insensitive to realistic variations of the thermal
prescription, at sub-percent accuracy.  We report on the thermal
properties and ejecta of mergers with masses just above the threshold
mass, i.e., which experience a single core-bounce before
collapsing. During the bounce, the thermal pressure can reach
$\mathcal{O}(1-10)$\% of the cold pressure at supranuclear densities,
depending on the thermal treatment, leading to modest differences in the dynamical ejecta that are launched and in the remnant disk mass as a result. 
\end{abstract}

\maketitle

\section{Introduction}
 
Binary neutron star (BNS) mergers provide an exciting laboratory for
studying dense-matter physics. The first BNS merger, GW170817, was
observed both from its inspiral gravitational waves (GWs)
\cite{LIGOScientific:2017vwq}, as well as from the electromagnetic
(E/M) counterpart to the event \cite{LIGOScientific:2017ync}.  The
inspiral GWs led to the constraints on the neutron star
tidal deformability, which enabled novel constraints on the
still-uncertain dense-matter equation of state (EoS)
\cite{LIGOScientific:2017vwq,LIGOScientific:2018hze,LIGOScientific:2018cki}
\cite[for reviews,][]{Baiotti:2016qnr,Raithel:2019uzi,Baiotti:2019sew,
  Chatziioannou:2020pqz,Radice:2020ddv,Chatziioannou:2024tjq}.  To
date, a second BNS merger and several neutron star-black hole mergers
have also been observed, via their inspiral GWs
\cite{LIGOScientific:2020aai,LIGOScientific:2021qlt,LIGOScientific:2024elc}.
However, none have had sufficient signal-to-noise ratio (SNR) to
provide informative constraints on the tidal deformability, or thus
new constraints on the EoS \cite[e.g.,][]{Chatziioannou:2024tjq}.

The fate of the remnant provides another important probe of the EoS --
one that is complementary to constraints from the tidal deformability
and that may be possible even for events with lower GW SNR, if the
event is accompanied by an E/M counterpart, see e.g.,
\cite{Bauswein:2017vtn,Radice:2017lry,Kiuchi:2019lls}.

In particular, following a BNS merger, there are two possible
outcomes: the remnant object can collapse promptly to a black hole, or
it can survive -- at least temporarily -- as a massive neutron star
remnant.  In the latter scenario, the remnant may undergo a delayed
collapse on a secular timescale or survive indefinitely, depending on
the mass of the binary and the EoS
\cite[e.g.,][]{Paschalidis:2016vmz,Baiotti:2016qnr,Radice:2020ddv}.
The mass above which a remnant collapses promptly to a black hole is
called the threshold mass for prompt collapse, \Mthresh.  The term
``prompt collapse" is typically taken to mean that there is no core
bounce \cite{Shibata:2005ss,
  Hotokezaka:2011dh,Bauswein:2013jpa,Kiuchi:2019lls,Kashyap:2021wzs}.
This is equivalent to requiring the maximum rest-mass density to
monotonically increase after merger, or the minimum lapse to
monotonically decrease. Another definition of \Mthresh, based on the
formation time for the black hole after merger, gives qualitatively
similar results \cite{Koppel:2019pys,Tootle:2021umi} \cite[but see
  also][]{Ecker:2024kzs}. 
  
However, ``prompt collapse"  -- defined
  in the sense that the black hole forms within a few dynamical/free-fall
  timescales following first contact, i.e. $t_d\simeq
2\sqrt{R^3/M}\simeq 0.15 {\rm ms} (R/14 \rm{km})^{3/2}(M/2.8M_\odot)^{-1/2}$, where $R$ and
  $M$ are the radius and mass of the remnant neutron star -- 
  can also occur even
  if there is a single core-bounce prior to
  collapsing~\cite{Kiuchi:2019lls}. This scenario is sometimes
  referred to as bounce-collapse. The bounce-collapse case is
  qualitatively different from the scenario in which the matter has enough
  angular momentum to form a transient hypermassive neutron star that
  collapses on a GW timescale, the Alfv\'en timescale, or  
  even the much longer cooling timescale 
  ~\cite[e.g.,][]{Sekiguchi:2011zd,Paschalidis:2012ff}.

Early BNS simulations showed that the threshold mass for prompt collapse typically exceeds
the maximum mass of a cold, non-rotating neutron star ($M_{\rm
  max}^{\rm TOV}$) by $\sim$30-70\%, according to
  \begin{equation}
 M_{\rm thresh} = k M_{\rm max}^{\rm TOV},
 \end{equation}  
with softer EoSs leading to smaller $k$ and stiffer
EoSs larger $k$ \cite{Shibata:2005ss,Shibata:2006nm,
Hotokezaka:2011dh,Bauswein:2013jpa}.
Various studies since have found that $k$ can be 
parametrized in terms of the compactness
of the maximum-mass neutron star 
\cite[e.g.,][]{Bauswein:2013jpa,Koppel:2019pys,Kashyap:2021wzs},
tidal deformability near the maximum mass \cite{Bauswein:2020aag},
or the incompressibility of cold nuclear matter
\cite{Perego:2021mkd},
with small corrections 
depending on the mass ratio
\cite{Bauswein:2020xlt,Kolsch:2021lub} or spins 
of the initial neutron stars \cite{Tootle:2021umi,Schianchi:2024vvi}.
These dependencies can be summarized
 as a quasi-universal relation for 
 \Mthresh($Q_{\rm EoS}$, \Mtov),
where ``$Q_{\rm EoS}$"  refers generically to
a quantity that depends on the EoS.

In the case of GW170817, the prompt collapse outcome is generally
disfavored because GW170817 was accompanied by a bright
kilonova~\cite{Bauswein:2017vtn,Margalit:2017dij,Radice:2017lry}
\cite[but see][]{Kiuchi:2019lls}, implying that the binary mass of
GW170817 must have been below \Mthresh.  Together with the
quasi-universal relations for \Mthresh, this constraint has been used
to place lower limits on the neutron star radius
\cite{Bauswein:2019ybt,Bauswein:2017vtn,Koppel:2019pys}.
  
Such constraints are inherently limited by the scatter in the
\Mthresh~universal relationship, which assumes that (for a fixed mass
ratio and spin) \Mthresh~is only a function of \Mtov and the radius,
tidal deformability, or nuclear incompressibility.  One complication
arises due to the fact that these are all properties of the
\textit{cold} (i.e., zero-temperature) EoS.  During a core bounce,
however, shocks can heat the remnant and potentially provide
additional thermal support that delays the collapse, or the shocks can
redistribute angular momentum to give rise to larger disks around the
remnant black hole.  Thus, \Mthresh~ and the disk mass onto the
remnant black hole may also depend on the finite-temperature part of
the EoS.

One of the goals of this work is to investigate the sensitivity of
\Mthresh~to finite-temperature effects, in order to better quantify
the uncertainties in this quasi-universal relation. To that end, we
adopt a parametric approach, in which the EoS pressure is
decomposed into a cold and thermal component, according to
\begin{equation}
P_{\rm total} = P_{\rm cold} + P_{\rm th}.
\end{equation}
This construction is a common approach 
for extending cold EoSs to finite temperatures in numerical simulations.
Traditionally, the thermal pressure $P_{\rm th}$ has been approximated
as an ideal gas, with a constant thermal index $\Gamma_{\rm th}$
\cite{Janka1993,Bauswein:2010dn}. Using
this constant-$\Gamma_{\rm th}$ approach (also called the ``hybrid approach"),
Ref.~\cite{Bauswein:2020xlt} found that \Mthresh~
can be sensitive to the choice of the thermal index
 \cite[see also][]{Blacker:2023afl}, suggesting that
 thermal effects indeed play a role in determining the outcome of 
 a BNS merger near the threshold for prompt collapse.

However, the ideal-fluid approximation of the hybrid approach neglects
the density-dependence of $\Gamma_{\rm th}$, which is significant in
the degenerate-matter cores of neutron stars
\cite{Constantinou:2015mna,Carbone:2019pkr,Huth:2020ozf}.  In order to
go beyond this simplified model, Ref.~\cite{Raithel:2019gws}
introduced a new parametric framework for calculating the
thermal pressure, that includes the leading-order effects of
degeneracy at high densities.  The framework is based on Landau's
Fermi Liquid Theory, in which the thermal contribution to the pressure
depends only on the particle effective mass function $M^*(n)$
\cite{Baym1991,Constantinou:2015zia}.  The framework of
Ref.~\cite{Raithel:2019gws} parametrizes the thermal pressure in terms
of an accurate, two-parameter model for $M^*(n)$, thus providing a
method for realistically and generically extending any cold nuclear
EoS to finite temperatures. This approach has been shown to accurately
capture the density- and temperature-dependence of realistic
finite-temperature EoS tables both analytically
\cite{Raithel:2019gws}, as well as in merger simulations in full
general relativity \cite{Raithel:2022nab}.
Other works have since developed additional 
phenomenological frameworks for constructing EoSs with  
 density-dependent thermal effects, based on a liquid
 drop model with Skyrme interactions
 \cite{Fields:2023bhs,Jacobi:2023olu}. These provide
 another approach for modeling finite-temperature
 EoSs in BNS simulations (for a comparison of the former
 of these frameworks
 with the $M^*$-approach, see Appendix B of \cite{Raithel:2023zml}).

In this work, we explore the sensitivity of the threshold mass
to thermal effects, using the $M^*$-framework of
\cite{Raithel:2019gws}. We investigate the threshold mass 
for one soft and one stiff cold EoS, which span a wide range 
of stellar compactness. For each EoS, we determine 
the threshold mass by simulating
a series of BNS mergers of varying masses in full general relativity,
with different choices for the thermal prescription.
We then explore in detail the
dynamics, thermal evolution, and ejecta
of the highest-mass binary that undergoes a single
core-bounce during collapse (i.e., a bounce-collapse), for each EoS.

We find that the threshold mass for prompt collapse is practically
insensitive to the details of the thermal prescription, for the wide
range of $M^*$-parameters explored in this work bracketing the
expected possibilities from realistic finite-temperature equations of
state. When exploring the evolutions for the
bounce-collapse binaries, we find small differences in the thermal
profiles of the merging cores. 
Although these thermal differences
do not affect the threshold mass, we find
that they lead to modest differences
($\lesssim40\%$) in the ejecta that are launched during
the bounce-collapse evolutions and  
in the bound disk masses that form around the remnant black hole
for most of the thermal treatments explored here. We
find an exception for the case of the stiff cold EoS and 
one particular thermal treatment, for which
there is a suppressed core bounce that
is associated with a $\sim60\%$ increase in the resulting
dynamical ejecta, illustrating that finite-temperature
uncertainties can still impact some observable outcomes,
even if the threshold mass is not affected. Our findings can potentially affect constraints on the tidal deformability set by arguments based on the disk mass/ejecta arising in prompt collapse mergers, and motivate further the study of unequal mass prompt collapse BNS mergers. 

We start in Sec. \ref{sec:methods} with an overview of our numerical
methods.  We describe our overall results in Sec.~\ref{sec:results},
with a detailed discussion of the results for the soft cold EoS in
Sec.~\ref{sec:results_soft} and for the stiff cold EoS in
Sec.~\ref{sec:results_stiff}. We discuss implications of these
findings, in particular in the context of constraints from GW170817,
in Sec.~\ref{sec:discussion}.  Throughout this work, we use natural
units of $G=c=1$.  We also note that, unless otherwise specified,
binary masses (including \Mthresh) in this work refer to the total
gravitational mass of the binary at infinite separation.

\begin{table}
\caption{Summary of the $M^*$ parameters ($n_0$ and $\alpha$) explored in this work.}
  \label{table:Mstar}
\centering
\begin{tabular}{ccccc}
\hline \hline
 $M^*$-parameters & Thermal Case \\
 \hline
     (0.08~fm$^{-3}$, 0.6) &   I  \\
   (0.08~fm$^{-3}$, 1.3) &   II  \\
       (0.22~fm$^{-3}$, 0.6) &   III  \\
       (0.22~fm$^{-3}$,  1.3) &  IV  \\
\hline \hline
\end{tabular}
\end{table}

\section{Numerical methods}
\label{sec:methods}
In this section, we describe the numerical methods used
in this work, including the construction of
the finite-temperature EoSs, the process for determining
the threshold mass for prompt collapse, and the set-up
of our general relativistic BNS merger simulations.

\subsection{Equations of state}
\label{sec:EOS}
In order to investigate the sensitivity of \Mthresh~ to thermal effects, 
we explore two zero-temperature EoSs that span a wide range of
 stellar compactness. For each of these cold EoSs, we add
on one of four different thermal extensions, following the 
framework of \cite{Raithel:2019gws}.

\begin{figure}[ht]
\centering
\includegraphics[width=0.4\textwidth]{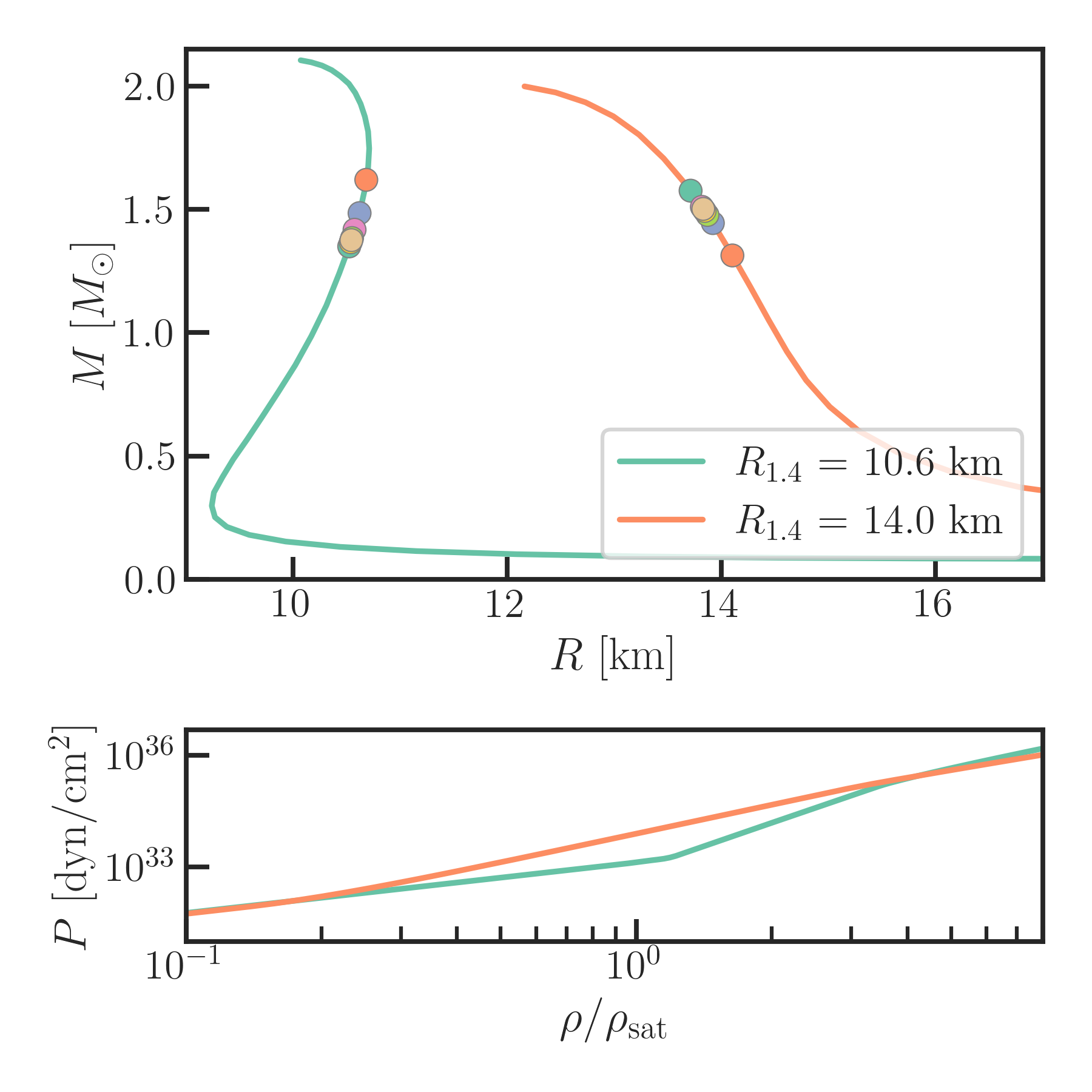}
\caption{\label{fig:MR} Top: Mass-radius curves for the two EoSs
 considered in this work. The markers indicate the masses included
  in our series of equal-mass BNS merger simulations. Bottom: 
  Pressure as a function of density, relative to the nuclear saturation 
  density ($\rns=2.7 \times 10^{14}$g/cm$^3$), for the same models.}
\end{figure}

\begin{table*}
\caption{Series of simulated binary masses, for the set of soft and stiff EoSs considered
in this work. $M_{g, \infty}$ indicates the total gravitational mass at infinite separation,
while $M_{\rm ADM}$ represents the corresponding Arnowitt-Deser-Misner mass
for a system with the same baryon mass, at the initial separation. $M_{b}$ denotes the total baryon mass.
The final column indicates the fate of the remnant object, 
for simulations with Thermal Treatment I.}
 \label{table:masses}

\centering
\begin{tabularx}{0.95\textwidth}{ c@{\hskip 0.5in}c@{\hskip 0.5in}c@{\hskip 0.5in}c@{\hskip 0.5in}c@{\hskip 0.5in}c@{\hskip 0.5in}}
\hline \hline
 EoS  & Description & $M_{g,\infty} (\Ms) $ & $M_{\rm ADM} (\Ms) $ & $M_{b} (\Ms)$  & Outcome
   \\ 
 \hline 
 		& Initial guess 	& 2.70 &  2.67 & 3.04 & Stable past 1st bounce  \\
		& $1.2\times$ larger & 3.24 &  3.20 & 3.74 & Prompt collapse \\
		& Step 1 		& 2.97 &  2.93  & 3.38 & Prompt collapse\\
Soft		& Step 2 		& 2.84 &  2.80  & 3.21 & Prompt collapse \\
		& Step 3 		& 2.77 &  2.73 & 3.12 & Prompt collapse  \\
		& Step 4 		& 2.73 &  2.70 & 3.08 & Bounce collapse \\
		& Step 5 		& 2.75 &  2.72 & 3.10 & Bounce collapse \\
\hline
 		& Initial guess 	& 3.15 &  3.14 & 3.48 & Prompt collapse \\
		& $1.2\times$ smaller & 2.63 &  2.62 & 2.85 &  Stable past 1st bounce \\
		& Step 1 		& 2.89 &  2.88  & 3.16 & Stable past 1st bounce \\
Stiff		& Step 2 		& 3.02 &  3.01  & 3.32 & Prompt collapse \\
		& Step 3 		& 2.96 &  2.95  & 3.24 & Bounce collapse \\
		& Step 4 		& 2.99 &  2.98 & 3.28 & Bounce collapse \\
		& Step 5 		& 3.01 &  2.99  & 3.30 & Prompt collapse \\

\hline
\end{tabularx}
\end{table*}

For the zero-temperature models, we adopt one soft and one stiff EoS,
to span a wide range of stellar compactness. The soft EoS 
was chosen to match the softest EoS studied in \cite{Kiuchi:2019lls}
that was able to explain the observed luminosity of the 
E/M counterpart to GW170817 \cite{LIGOScientific:2017ync}. This corresponds
to the model with binary tidal deformability $\widetilde{\Lambda}=242$
for mass ratio $q=0.774$ in that work. We re-parameterize this EoS
with smoothed piecewise polytropes, following
\cite{OBoyle:2020qvf,Raithel:2022san} to ensure continuity and
differentiability in the pressure at all densities.
For details of this fit, see Appendix~\ref{sec:appendixEOS}.
The resulting cold EoS predicts the radius of a 1.4~$\Ms$
neutron star to be $R_{1.4}$=10.6~km  and 
\Mtov$ = 2.11 \Ms$. 

For the stiff cold EoS, we use a parametrized version of
 the H4 tabular EoS \cite{Lackey:2005tk}, fit 
 with the same smoothed piecewise polytropic framework 
 \cite{OBoyle:2020qvf, Raithel:2022san}. This EoS has been 
 simulated previously in \cite{Raithel:2023zml}. It predicts 
 $R_{1.4}$=14.0~km and \Mtov$=2.01~\Ms$.
 We show both EoSs and their mass-radius relations
 in Fig.~\ref{fig:MR}. These models thus bracket current astrophysical constraints on neutron star radii and the maximum mass \cite[e.g.,][]{Chatziioannou:2024tjq}.

We extend these cold EoSs to finite temperatures using
the $M^*$-framework of \cite{Raithel:2019gws}. The framework
is based on a two-parameter model of the particle effective mass 
function, $M^*(n)$. The free parameters of this model include
a density parameter $n_0$, which controls the density
above which $M^*(n)$ starts to decrease away from its
rest-mass value, and a power-law parameter $\alpha$, 
which controls the rate at which $M^*(n)$ decreases.
These parameters can be roughly interpreted as the density
at which particle-interactions start to become significant
and the relative strength of those interactions, respectively. 
The M$^*$-framework additionally includes
the thermal contribution of leptons 
\cite[for details, see][]{Raithel:2019gws,Raithel:2021hye}.

In this work, we construct four different thermal
prescriptions, corresponding to four different choices
of $M^*$-parameters, with $n_0 \in (0.08,0.22)$~fm$^{-3}$ and
$\alpha \in (0.6,1.3)$. This range of parameters
was previously found to bracket
 the range of uncertainty in realistic calculations
 of the effective mass function for commonly-used
 finite-temperature EoS tables  \cite{Raithel:2019gws}
 (see also Appendix B of \cite{Raithel:2023zml}
 for additional discussion). We refer to
 these parameter sets at Thermal Cases I-IV in the following, and
 summarize the parameters in Table~\ref{table:Mstar}.

Finally, in extending the cold EoSs to finite-temperatures, we follow
Refs. \cite{Raithel:2021hye,Raithel:2023zml} in assuming that the
matter remains in its initial (cold) $\beta$-equilibrium composition during
the inspiral and through the merger. To impose these conditions, we use the cold EoS to set the
leading-order coefficients of the nuclear symmetry energy to be
$S_0$=31~MeV and $L$=19.3 for the \Rsmall cold EoS, and $S=32$ and
$L=$112.1~MeV for the \Rbig EoS (for details, see
\cite{Raithel:2019gws,Raithel:2021hye,Raithel:2023zml}); this procedure assumes that the neutrinos escape and instantaneously restore
the zero-temperature $\beta$-equilibrium composition at each 
time-step. This is a simplification, and one of the extreme 
ends one can assume (the other being that neutrinos do not 
have time to escape, in which case the electron fraction can 
then be advected). While the latter choice is likely more realistic for determining the composition, the present choice allows us to cleanly
separate the impact of thermal and out-of-equilibrium effects, 
and to focus in this work on the former.
To quantify the effect this may have on our work,
Appendix~\ref{sec:composition} estimates the 
fractional change to the total pressure
introduced by neglecting the composition,
for the temperatures and densities
probed in our simulations. For some minimal assumptions, we
estimate that composition effects contribute
$\lesssim2\%$ fractional change to the total pressure,
beyond what is already captured with our thermal-only model.
We will investigate different treatments of neutrinos and
out-of-equilibrium effects in a future study, 
and stress that such an 
investigation is important for understanding
the complete picture of EoS effects on the threshold mass,
but these estimates motivate focusing on thermal
effects as a first step towards solving the full problem.

\subsection{Determination of the threshold mass}
In order to determine the threshold mass for prompt collapse, 
we simulate an ensemble of BNS mergers at different total masses, 
but otherwise identical initial conditions (described below). 
To determine the masses to simulate, we use a bisection
 search, starting from the value for \Mthresh($Q_{\rm EoS}$,\Mtov)
 predicted by the quasi-universal relations of
 \cite{Bauswein:2020xlt} for each of our cold EoSs.
  We use ``Fit no. 1" of that work, which was calibrated to a 
 large set of BNS simulations and which
 depends on the radius of a 1.6~$\Ms$ neutron star
 (such that $Q_{\rm EoS} = R_{1.6}$) 
 and
 \Mtov. As in Ref.~ \cite{Bauswein:2020xlt}, 
 all binary masses in this work refer to the total
 gravitational mass of the binary at infinite orbital
 separation, unless otherwise specified.
 
For each cold EoS, we thus construct a BNS system with the
 total binary mass predicted from this relation 
 and evolve the system through merger.  The formation of
 a black hole is indicated by the appearance of an apparent horizon,
  which we track using the module \texttt{AHFinderDirect}
 \cite{Thornburg:1995cp,Thornburg:2003sf} 
  
 If the remnant collapses promptly to a black hole --
 indicated by a monotonically decreasing minimum lapse --
 we reduce the binary mass by
 20\% for the second evolution in our series. Otherwise, we 
 increase the mass by 20\%. For both cold EoSs considered,
 this was a sufficiently large change in the mass to produce
 the opposite outcome (e.g., a no-bounce collapse, if the first
 remnant was initially stable). We then perform a bisection 
 search with five additional simulations, to determine the 
 threshold mass to an accuracy 
 of $0.2 / 2^5 = 0. 625\%$. For the final determination,
 we average together the masses of the two binaries
  that are closest to the threshold mass,  i.e., 
 \begin{equation}
 M_{\rm thresh}=\frac{1}{2} \left( M_{\rm sub} + M_{\rm sup} \right)
 \end{equation}
 where $M_{\rm sub}$ is the subcritical mass (leading to a single core
 bounce collapse) and $M_{\rm sup}$ is the supercritical mass (leading
 to a no-bounce collapse).
 
 We report the binary masses from the bisection 
 searches for both cold EoSs, evolved with Thermal 
 Treatment I, in Table~\ref{table:masses}.
For reference, Table~\ref{table:masses}
also provides the Arnowitt-Deser-Misner (ADM) mass 
for each binary, corresponding to a system with the same
baryon mass at the initial binary separation.

\subsection{Initial conditions and numerical set-up}

The simulations in this work were performed using the 
dynamical spacetime, general-relativistic (magneto)-hydrodynamics
code of \coderefs. The code is built within the Cactus/Carpet
framework \cite{Allen2001,Schnetter:2003rb,Schnetter:2006pg},
and was recently extended in \cite{Raithel:2021hye,Raithel:2022san}.
We refer the reader to these works for detailed information
on the code.

For each simulation in our bisection series, we construct initial data using 
\texttt{LORENE} \cite{Lorene}.  In each case, the initial configuration
describes two equal-mass, irrotational, unmagnetized, neutron 
stars at zero-temperature in a quasi-circular orbit, with initial
separation of 32.5 km for the soft EoS binaries and 40 km for the
 stiff EoS binaries.

Our simulations use nine spatial refinement levels, separated by a 2:1 
refinement ratio. The grid is set up such that the innermost refinement 
level is approximately 30\% larger than the coordinate diameter of the 
initial neutron stars. The resulting computational domain corresponds 
to a box of radius 2880~km (3900~km) for the soft (stiff) EoS models. 
We impose equatorial symmetry to reduce computational cost.

The finest-level grid spacing is $\Delta x=$ 140.625~m for the soft EoS binaries
and $\Delta x$= 195~m for the stiff EoS binaries. This provides the same 
effective resolution for both EoSs, with $\sim115$ grid points covering
the coordinate diameter of each initial neutron star, along the x-direction. 

For numerical stability and to ensure that velocities do not become unphysical, especially in cases where matter catastrophically collapses to form a black hole, numerical relativity codes impose a maximum allowed Lorentz factor ($\gamma$) for the fluid (as measured by a normal observer), see e.g.~\cite{Chang:2020ktl}. In our case we typically choose $\gamma_{\rm max}=50-100$. However, in some evolutions with \Mtot~ near the threshold mass for prompt
collapse, we find that a more aggressive velocity limit (as aggressive as
$\gamma_{\rm max}=15$) is needed to continue the simulations past black hole
formation. In this work, we choose $\gamma_{\rm max}=100$ unless numerical stability requires $\gamma_{\rm max}=15$.
For the core-bounce
evolutions described in this work and for cases where we were able to use $\gamma_{\rm max}=15$ or $\gamma_{\rm max}=100$, we find that the choice of Lorentz factor cap does not affect the
outcome, dynamics, or thermal properties of the dense-matter core. In contrast, we find, modest differences in the dynamical ejecta and disk mass 
($\lesssim20\%$ and 23\%, respectively), as a result of changing the cap on the
Lorentz factor. We include some of our tests on the Lorentz factor limit and 
discuss the $\gamma_{\rm max}$
choices in more detail in Appendix \ref{sec:caps}.

Except where explicitly indicated above, the rest of the details of
the evolutions (equations and numerical methods) are identical to
those described in \cite{Raithel:2021hye,Raithel:2023zml}.

\section{Results}
\label{sec:results}

Here we describe the outcomes of our study. Our main conclusion is that \Mthresh~
does not depend on the finite-temperature part of the EoS. However, the finite-temperature part of the EoS can be responsible for changes in dynamical ejecta and the disk mass onto the remnant black hole.
We find this to be true for both cold EoSs explored here. In the following,
we describe this result in detail and explore the 
thermal profiles of the remnants and the 
ejecta from simulations just above the threshold mass,
which experience a single core bounce.
This provides a first look at how the thermal profiles,
dynamical ejecta, and disk masses from
a bounce-collapse merger can vary, 
for realistic variations of the thermal prescription of the nuclear EoS.

\subsection{Threshold mass for the \Rsmall cold EoS}
 \label{sec:results_soft}

We start with our results for the evolutions with the soft (\Rsmall) cold EoS.
To determine the mass threshold for prompt collapse, we
simulate BNS mergers with total masses
ranging from
\Mtot=2.70 to 3.24~$\Ms$, as summarized in Table~\ref{table:masses}. 
For the initial mass sequence, all binaries are evolved with Thermal Treatment I,
so that they are described by the same EoS (at both zero- and 
finite-temperatures) and differ only in their total mass.

\begin{figure}[ht]
\centering
\includegraphics[width=0.4\textwidth]{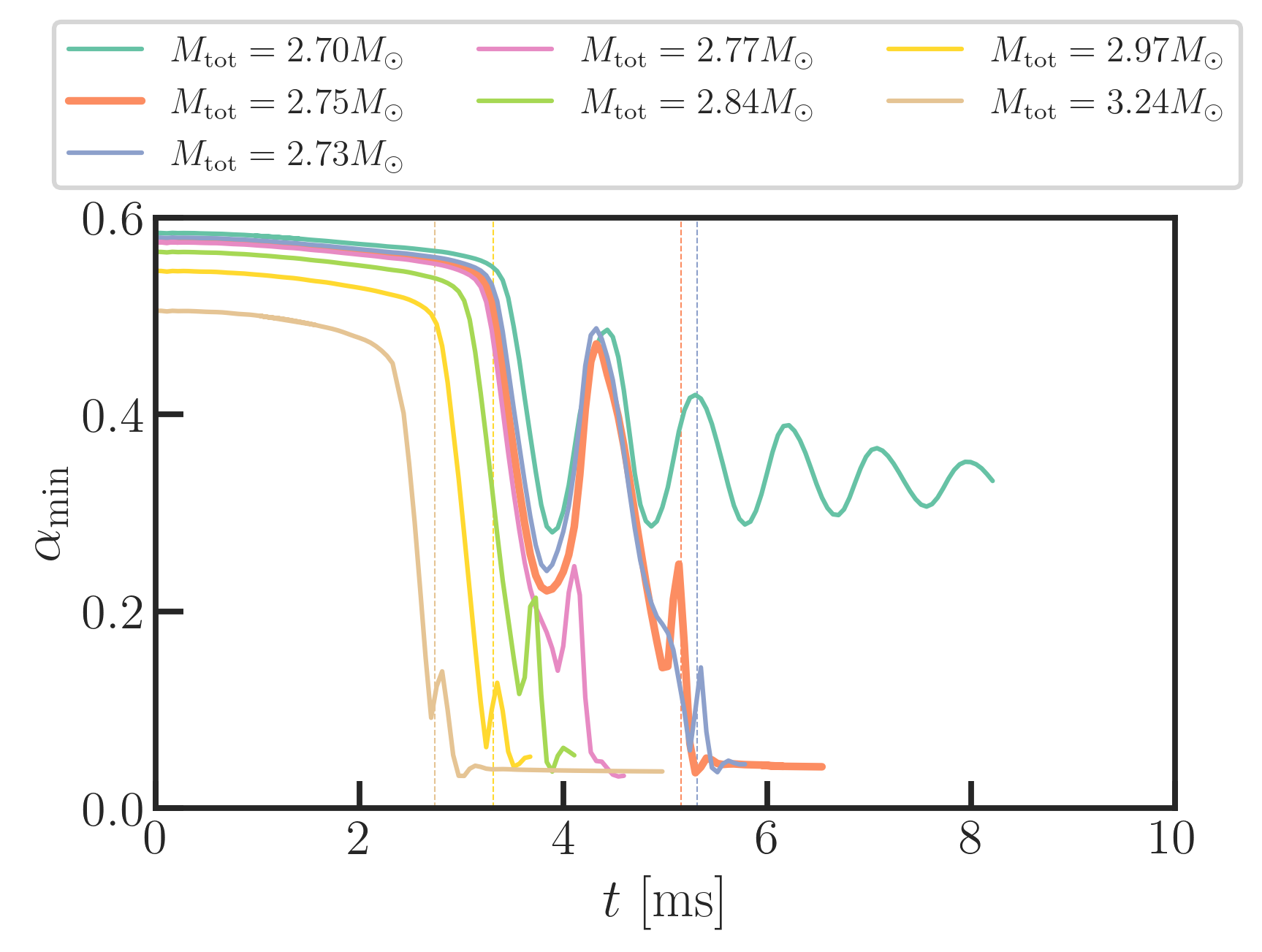}
\caption{\label{fig:lapse_allM}  Minimum lapse for the seven evolutions with the \Rsmall cold EoS and
the Case I thermal treatment. The colors indicate the binary mass for each evolution
(corresponding to the gravitational mass at infinite orbital separation).
The vertical dotted lines indicate the time at which the
apparent horizon is first found. The
largest mass that undergoes a single core bounce (\Msub=$2.75\Ms$)
 is emphasized here with the thick orange line.}
\end{figure}

\begin{figure}[ht]
\centering
\includegraphics[width=0.4\textwidth]{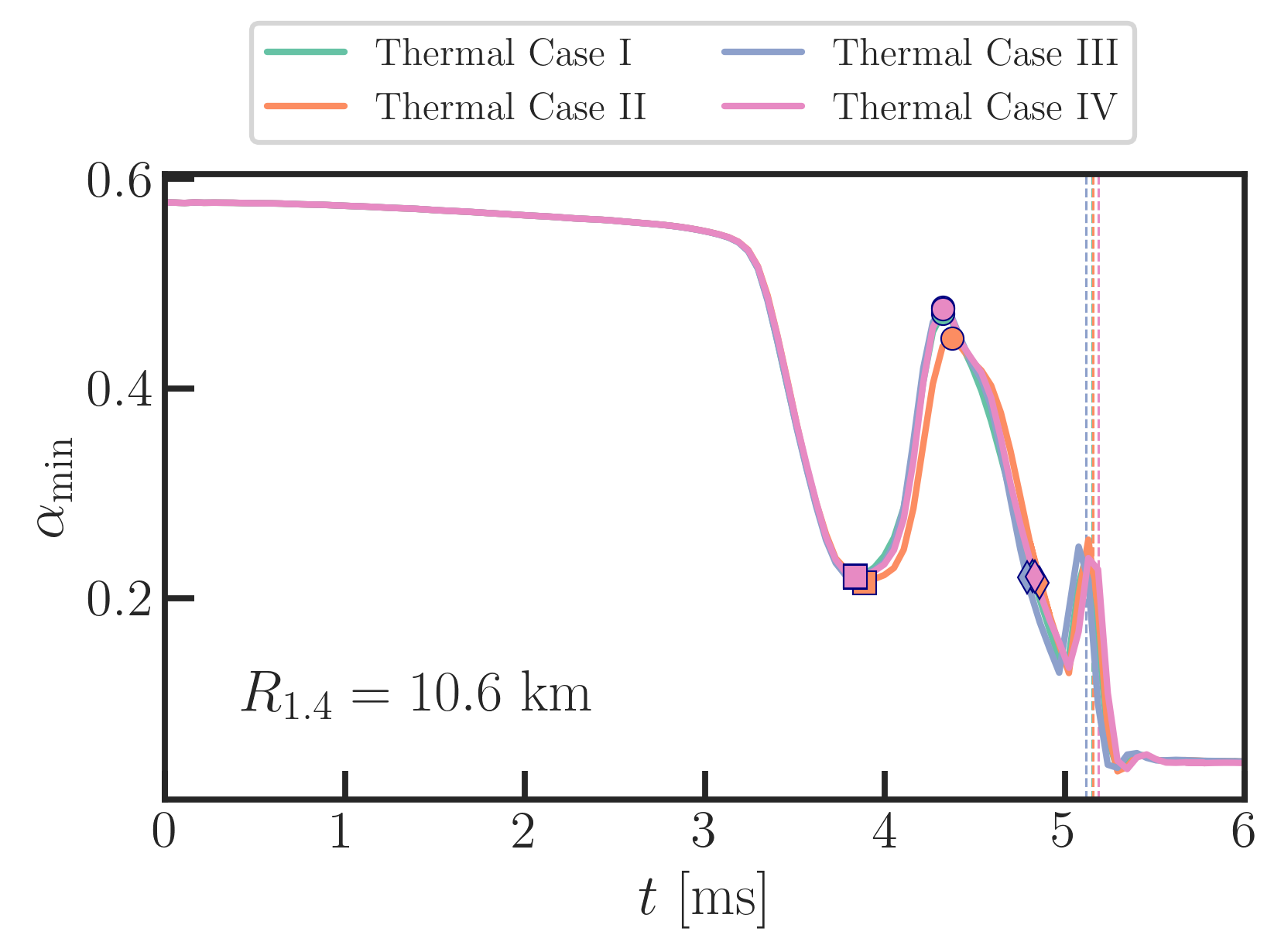}\\
\includegraphics[width=0.4\textwidth]{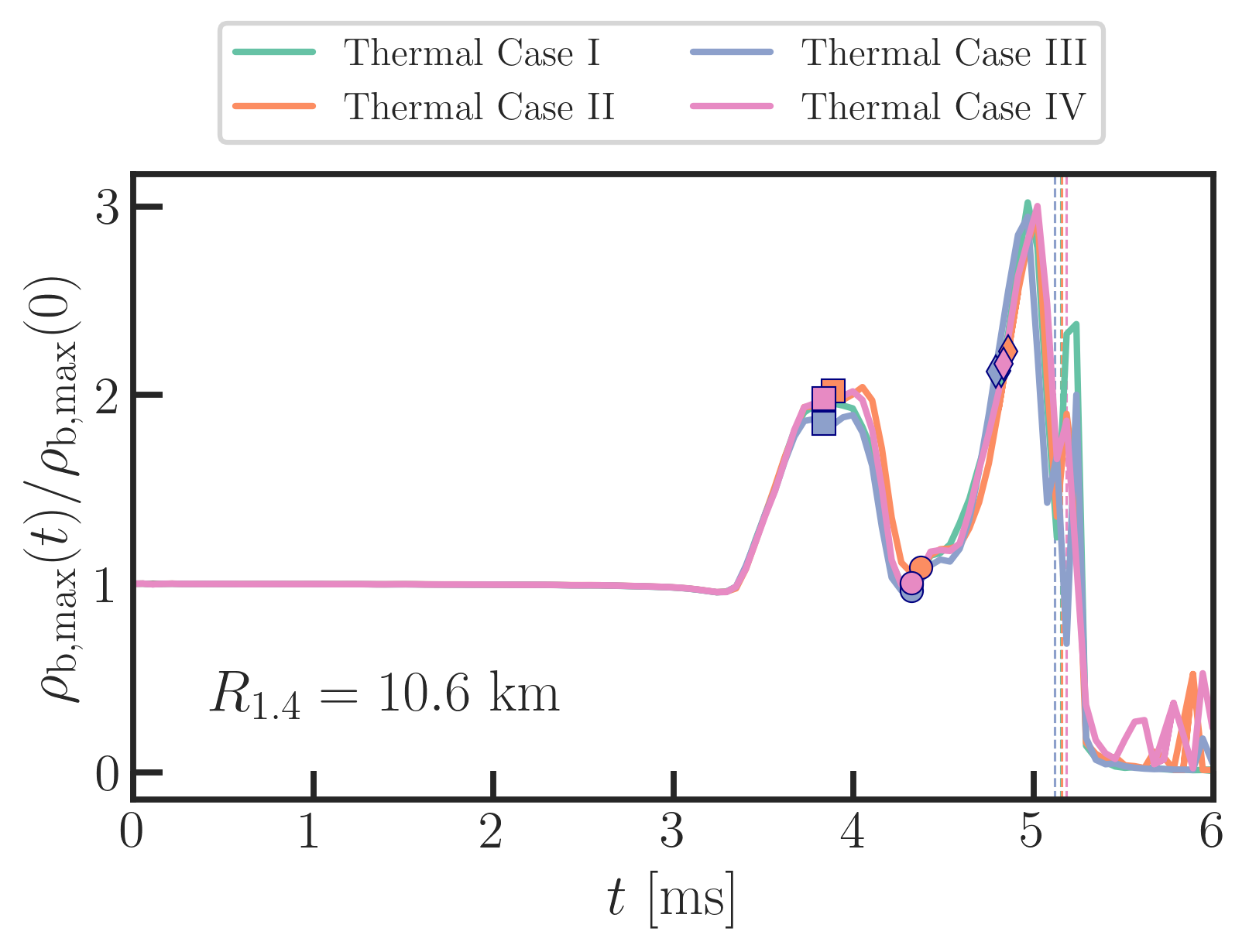}
\caption{\label{fig:lapse_Mthr_R10} \textit{Top:}  Minimum lapse for the bounce-collapse evolutions
(\Msub=2.75$\Ms$),  with the \Rsmall cold EoS and the four
different thermal treatments. The markers indicate the times of the 
the onset of the bounce, the end of the bounce, and 
one time on the way to collapse, for reference.
The vertical dotted lines indicate the time at which the
apparent horizon is first found.
\textit{Bottom:} Compression ratio for the same evolutions, showing the maximum rest-mass density relative to the initial rest-mass density.
}
\end{figure}

To track the progress of the evolution, we monitor the minimum value
of the lapse function, \amin, which we show in Fig.~\ref{fig:lapse_allM} for
all masses simulated.  The vertical dotted lines indicate the time at
which the apparent horizon is first found.

Figure~\ref{fig:lapse_allM} shows that all four evolutions with
$M_{\rm tot}\ge 2.77~\Ms$ result in a no-bounce collapse after
merger.  For the three evolutions with $M_{\rm tot}\le2.75~\Ms$, the
remnant undergoes at least one core bounce, before either collapsing
(for the more massive cases) or continuing to oscillate (as in the
less massive case). Averaging together these sub- and super-critical
masses, we thus conclude \Mthresh=2.76$\Ms$.

After determining \Mthresh~ in this way for Thermal Case I, 
we  repeat this procedure for the different thermal treatments.
To save on computational costs, we do not re-run the entire
 bisection sequence for each new thermal treatment. 
Instead, we simulate the two binaries bounding the critical mass,
$M_{\rm sub}$ and $M_{\rm sup}$. For all thermal treatments, 
we find that the two critical
masses result in the same set of outcomes  
-- i.e., \Msub~ leads to a bounce collapse, 
while \Msup~ leads to a prompt collapse --
thus confirming that the threshold mass is unchanged
by the details of the thermal treatment.

The top panel of Fig.~\ref{fig:lapse_Mthr_R10} shows the evolution of \amin~ for 
the bounce-collapse evolutions (i.e., with \Msub=2.75$\Ms$) 
for each of the four different thermal treatments, with the
\Rsmall cold EoS.  The filled squares in Fig.~\ref{fig:lapse_Mthr_R10}
 indicate the time of the first minimum
 in \amin~. This corresponds to onset of the bounce, $t_{\rm bounce}$,
 and is sometimes used to indicate the time-of-merger 
 \cite[e.g.,][]{Koppel:2019pys}. The filled circles
 in Fig.~\ref{fig:lapse_Mthr_R10} mark the
 end of the bounce, after which the remnant
 starts its collapse. Finally, the filled
 diamonds mark one additional time
 on the way to collapse, which we
 choose as 
 when \amin~ reaches the value of 
 its first minimum for the second time.
 
 We find negligible differences ($<0.1$~ms)
 in these three times, 
 between the different thermal treatments for this cold EoS.
 This is consistent with our previous work,
 which showed that the time-to-merger
 differs by $<0.1$~ms for these four thermal treatments, 
 albeit at a different mass and for a different
 cold EoS \cite{Raithel:2023zml}.

\begin{figure}[ht]
\centering
\includegraphics[width=0.4\textwidth]{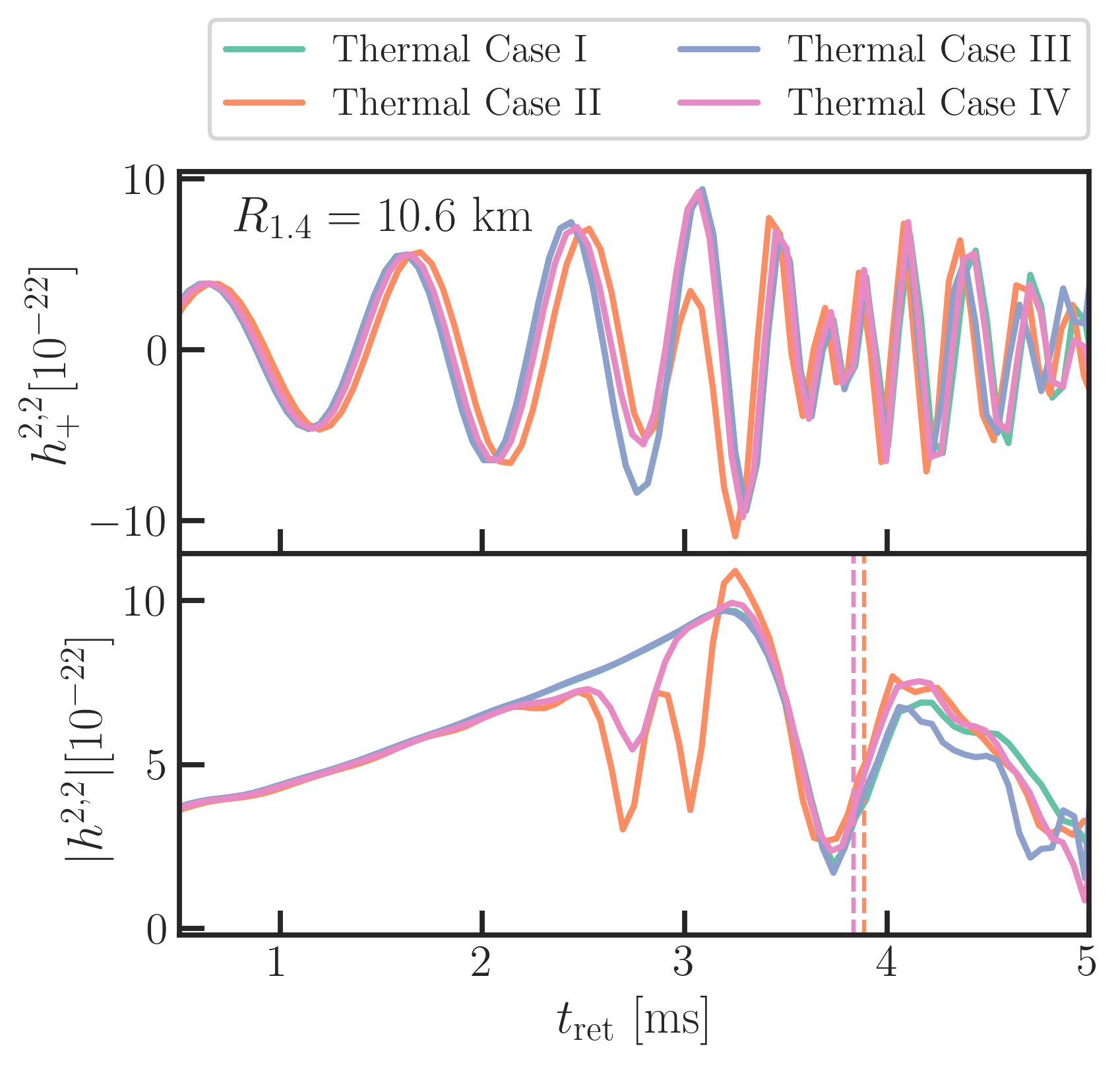}
\caption{\label{fig:gw_R10} Top: Plus-polarized gravitational wave strain for the $\ell=m=2$ mode from the bounce-collapse (\Msub) evolutions, with the \Rsmall cold EoS. Bottom: Total amplitude of the $\ell=m=2$ mode of the GW strain, $h=h_+ - i h_{\times}$. The strain is plotted relative to the retarded time, and has been scaled to a distance of 40 Mpc.
The vertical dashed lines indicate the coordinate time of the 
onset of the bounce (corresponding to the filled squares in Fig.~\ref{fig:lapse_Mthr_R10}).}
\end{figure}

  \begin{figure*}[ht]
\centering
\includegraphics[width=0.65\textwidth]{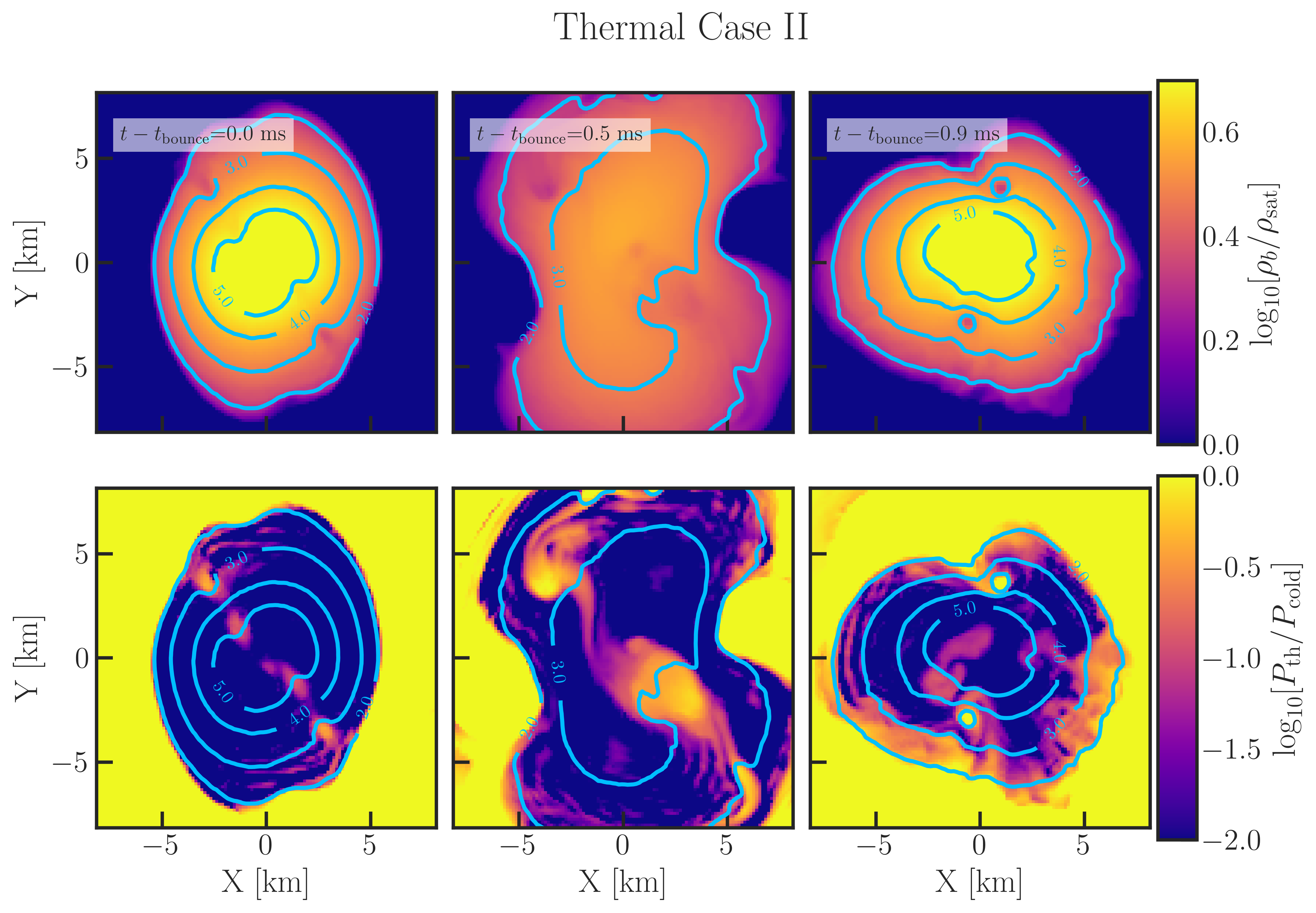}
\caption{\label{fig:Pth_R11_n08a13} 2D snapshots at times 
corresponding to the onset of the bounce, the end of the 
bounce, and one time on the way to collapse
for the \Msub~ evolution, evolved with the 
\Rsmall cold EoS and Thermal Case II. The top row shows the
 rest-mass density, $\rho_0$, relative to the nuclear saturation 
 density, $\rns$. The bottom row shows the thermal pressure, 
 relative to the cold pressure. In both
  rows, the blue lines indicate contours of constant rest-mass
   density, corresponding to 2, 3, 4, and 5$\times \rns$. }
\end{figure*}
 \begin{figure*}[ht]
\centering
\includegraphics[width=0.65\textwidth]{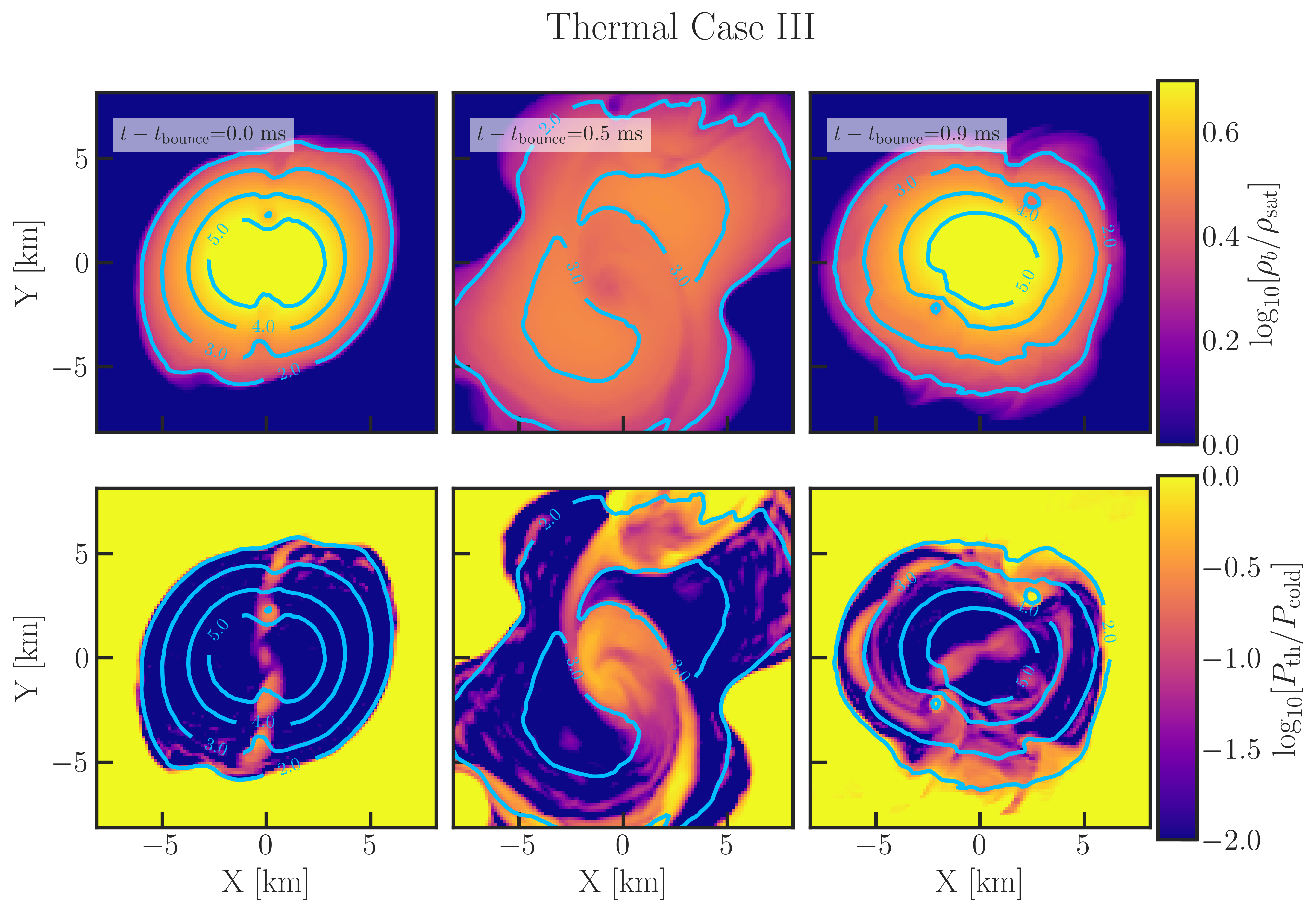}
\caption{\label{fig:Pth_R11_n22a06}  Same as Fig.~\ref{fig:Pth_R11_n08a13}, but with Thermal Case III, exhibiting a stronger thermal pressure component at the end of the bounce.}
\end{figure*}

We follow the literature in defining
the merger and other key times with respect to the minimum lapse
\cite[e.g.,][]{Koppel:2019pys,Tootle:2021umi}, but
we caution that the value of the lapse is a gauge-dependent quantity. A more robust definition of these times would come from the maximum rest-mass density and a gauge invariant definition would adopt the GWs. The bottom panel of Fig.~\ref{fig:lapse_Mthr_R10} shows the ratio of the maximum rest-mass density to the initial rest-mass density for the bounce-collapse evolutions. As in the top panel, the markers indicate the lapse-defined times of the onset of the bounce, the end of the bounce, and one point during collapse. We find that all four cases experience a compression ratio of $\sim2$ at the start of the bounce, and that the lapse-definitions of when the bounce starts and ends  indeed correspond roughly to local maxima and minima in the compression of the matter, as expected. In other words,
the start-of-the-bounce (as defined from the lapse) indeed corresponds to a significant compression of the matter, followed by a decrease in the compression ratio as the matter expands during the bounce.

For further confirmation,Fig.~\ref{fig:gw_R10} shows the $\ell=m=2$ mode of the plus-polarization of the GW strain, as well as the total amplitude of the $\ell=m=2$ mode of the strain, $h=h_+-i h_{\times}$. The dashed vertical line indicates the lapse-defined onset of the bounce, corresponding to the filled squares in Fig.~\ref{fig:lapse_Mthr_R10}. We find that this lapse-defined $t_{\rm bounce}$ is approximately associated with a local peak in the GWs, as expected
for the highly-compressed remnant. 
Thus, we find that the lapse-definition of $t_{\rm bounce}$ is consistent with these additional gauge-invariant diagnostics.

 \subsubsection{Thermal profiles for the bounce-collapse evolutions with the soft cold EoS} 
 \label{sec:thermal_soft}
 We now turn to a detailed look at the thermal properties of the
 bounce-collapse evolutions. To start, Fig.~\ref{fig:Pth_R11_n08a13} shows
 2D snapshots of the matter distribution for the bounce-collapse 
merger with the \Rsmall cold
EoS and Thermal Treatment II at the three different times
(start of bounce, end of bounce, and on the way to collapse)
that were indicated with markers in Fig.~\ref{fig:lapse_Mthr_R10}.
These snapshots correspond to equatorial slices through the XY plane.
The top row of Fig.~\ref{fig:Pth_R11_n08a13}
shows snapshots of the rest-mass density, relative to the nuclear saturation
 density $\rho_{\rm sat}=2.7\times 10^{14}$g/cm$^3$. 
 The bottom row shows the thermal pressure, plotted relative to the
 cold pressure. The blue lines in both rows indicate contours of constant 
 rest-mass density. For comparison, Fig.~\ref{fig:Pth_R11_n22a06}
 shows the same snapshots, but for the evolution with
  Thermal Treatment III. These two thermal cases
  are representative of the range of outcomes for this cold EoS.

\begin{figure*}[ht]
\centering
\includegraphics[width=0.95\textwidth]{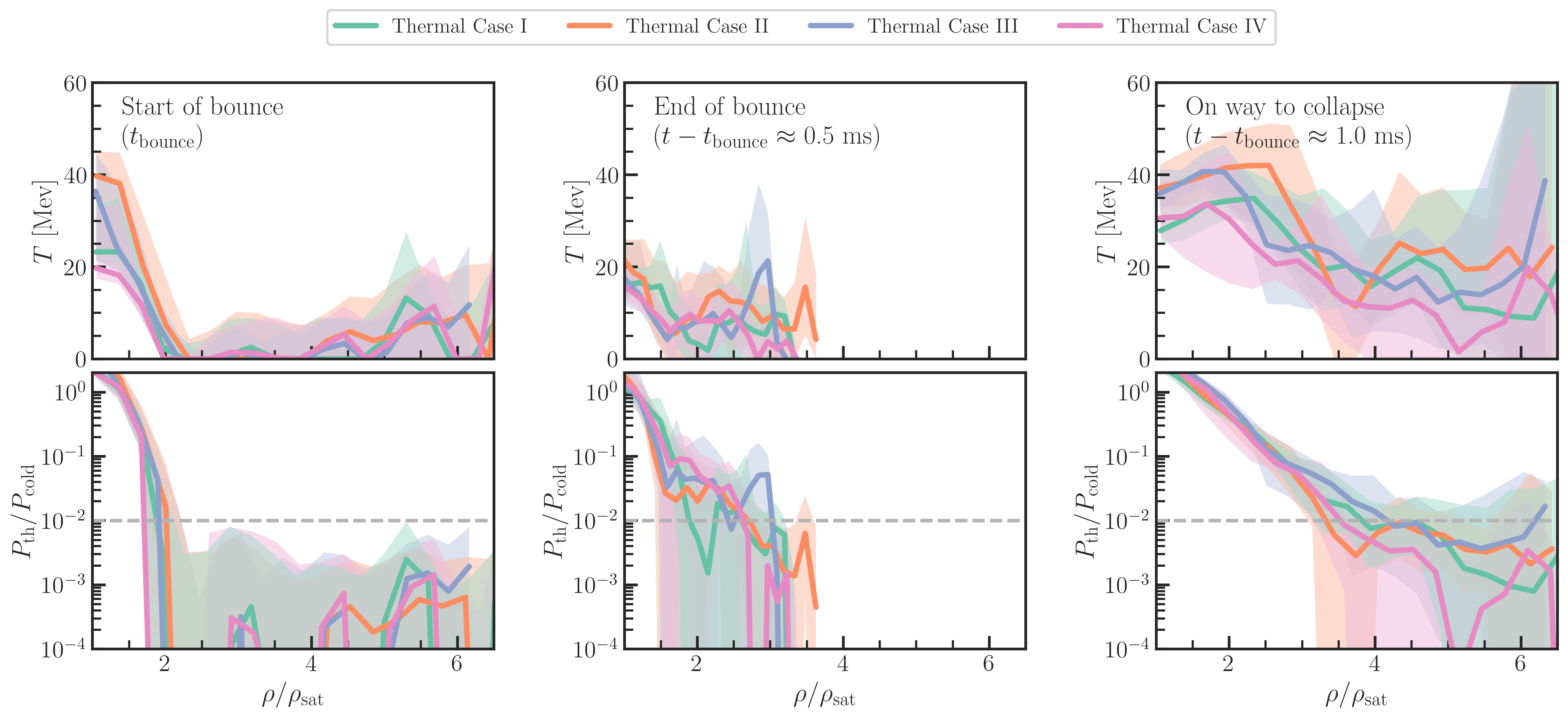}
\caption{\label{fig:medianPth}  Median thermal properties for the 
bounce-collapse (\Mtot=\Msub) evolutions, evolved with the
 \Rsmall cold EoS. From left to right, we show thermal quantities 
 extracted from 2D snapshots at the first minimum of the lapse
 (i.e., the start of the core bounce), the end of the bounce,
 and one point during the collapse
 (these times are indicated by the markers in Figs.~\ref{fig:lapse_Mthr_R10},
 and correspond to the same snapshots shown in
 Figs.~\ref{fig:Pth_R11_n08a13}-\ref{fig:Pth_R11_n22a06}). The
top row shows the median temperature, $T$, of all matter in 
each density bin. The bottom row shows the thermal pressure
 relative to the cold pressure, $P_{\rm th}/P_{\rm cold}$. The
 solid lines indicate the median values, calculated for bins
 that are spaced uniformly in density, with the shaded regions 
indicating 68\% bounds. }
\end{figure*}

\begin{table*}
\caption{Summary properties of the bounce-collapse (\Msub) evolutions.
 From left to right, the 
 columns indicate the cold EoS, the thermal prescription, the Lorentz factor limit
 (``s" for the standard cap, ``a" for the aggressive cap; see text) the average density (relative to the nuclear saturation density $\rns$), the density-weighted average of $P_{\rm th}/P_{\rm cold}$, the density-weighted average temperature, the (bound) disk mass, and the total amount of dynamical (unbound) ejecta. The averages are computed for matter at the end of the bounce, including all matter with densities  $\rho \ge \rho_{\rm sat}$, in order to focus on the dense-matter core.}
  \label{table:thermal}

\centering
\begin{tabular}{ccccccccc}
\hline \hline
Cold EoS && Thermal Case & Lorentz factor limit &  $\langle \rho_b/\rho_{\rm sat} \rangle$  &  $\langle P_{\rm th}/P_{\rm cold} \rangle$ & $\langle{T}\rangle$ [MeV] & $M_{\rm disk} (10^{-2} \Ms)$  & $M_{\rm ej} (10^{-3} \Ms)$ \\
\hline \hline
	      &~~&  I & s &  2.37  & 0.15 &  15.4 & 2.5 & 2.2   \\
\Rsmall  && II   &    a &  2.44  & 0.11 &   17.0 & 2.8 & 2.5	  \\
 	    &&  III & s  &  2.35  & 0.16 &    15.3 & 3.3 & 1.6 \\
	    &&  IV  & a  &  2.34  & 0.17  &   12.6 & 3.5 & 2.0 \\ \hline \hline
	     &&  I  &    s & 1.56  & 0.09 &   12.8 &  3.2 & 3.5 \\
\Rbig	    &&  II  & a  & 1.82  & 0.07 &    15.0 & 3.8 & 5.3 \\ 
	    &&  III & s & 1.53  & 0.09 &     10.3 & 3.7 & 4.0 \\
	    &&  IV  & s & 1.56  & 0.10  &      9.3 & 3.7 &  2.9	 \\\hline \hline
\end{tabular}
\end{table*}

 In both Figs.~\ref{fig:Pth_R11_n08a13} and ~\ref{fig:Pth_R11_n22a06}, 
 we see that at the start of the bounce, the matter remains 
 largely cold ($P_{\rm th} \ll P_{\rm cold}$), with small
 pockets of heating developing along the contact interface. However,
 there is significantly more heated material by the end of the bounce and
 during the subsequent collapse. The 2D snapshots in 
 Figs.~\ref{fig:Pth_R11_n08a13} and \ref{fig:Pth_R11_n22a06}
 indicate small differences in the heating depending
 on the thermal treatment used.
 
To better quantify these differences, we calculate the median 
value of $P_{\rm th}/P_{\rm cold}$, as well as the median temperature
 $T$, within uniformly-spaced density bins
for each 2D snapshot. The numerical grid included in these distributions
is exactly as shown in Figs.~\ref{fig:Pth_R11_n08a13} 
and \ref{fig:Pth_R11_n22a06}; that is, all 2D data in an equatorial slice through the XY plane, spanning the extent of the finest (innermost) refinement level.
We show these median quantities and 
1-$\sigma$ (68\%) bounds for all four thermal treatments
in Fig.~\ref{fig:medianPth}.

We find that at the onset of the bounce, all four thermal
treatments lead to negligible heating at densities above
$\sim2\rns$, with thermal pressures $\ll 0.01 P_{\rm cold}$.
At the end of the bounce, the median thermal pressure is
$\mathcal{O}$(1-10\%) of the cold pressure at
supranuclear densities. There is some variation
in the median $P_{\rm th}/P_{\rm cold}$
for the different thermal treatments, but
the 1-$\sigma$ intervals of all are overlapping,
indicating broadly similar thermal profiles.

\subsubsection{Ejecta from the bounce-collapse scenario}
\label{sec:ejecta_soft}

We additionally analyze the ejecta that is dynamically
launched during and immediately following the core bounce
for these evolutions.
We calculate the ejected mass by integrating
the total rest-mass density outside of a sphere of radius
$r=30M$, including all unbound matter ($-u^t >1$ and
$u^r>0$), according to
\begin{equation}
M_{\rm ej}(>r) = \int_{>r} \rho_b u^t \sqrt{-g} d^3 x
\end{equation}
where $u^t$ is the time-component of the four-velocity
 $u^r$ is the radial component, and $g$ is the determinant
 of the metric. We additionally calculate the disk mass, $M_{\rm disk}$, around the remnant black hole by integrating 
 the total rest-mass density of
 all \textit{bound} matter outside the apparent horizon, excluding the unbound (dynamical) ejecta.

 \begin{figure}[ht]
\centering
\includegraphics[width=0.45\textwidth]{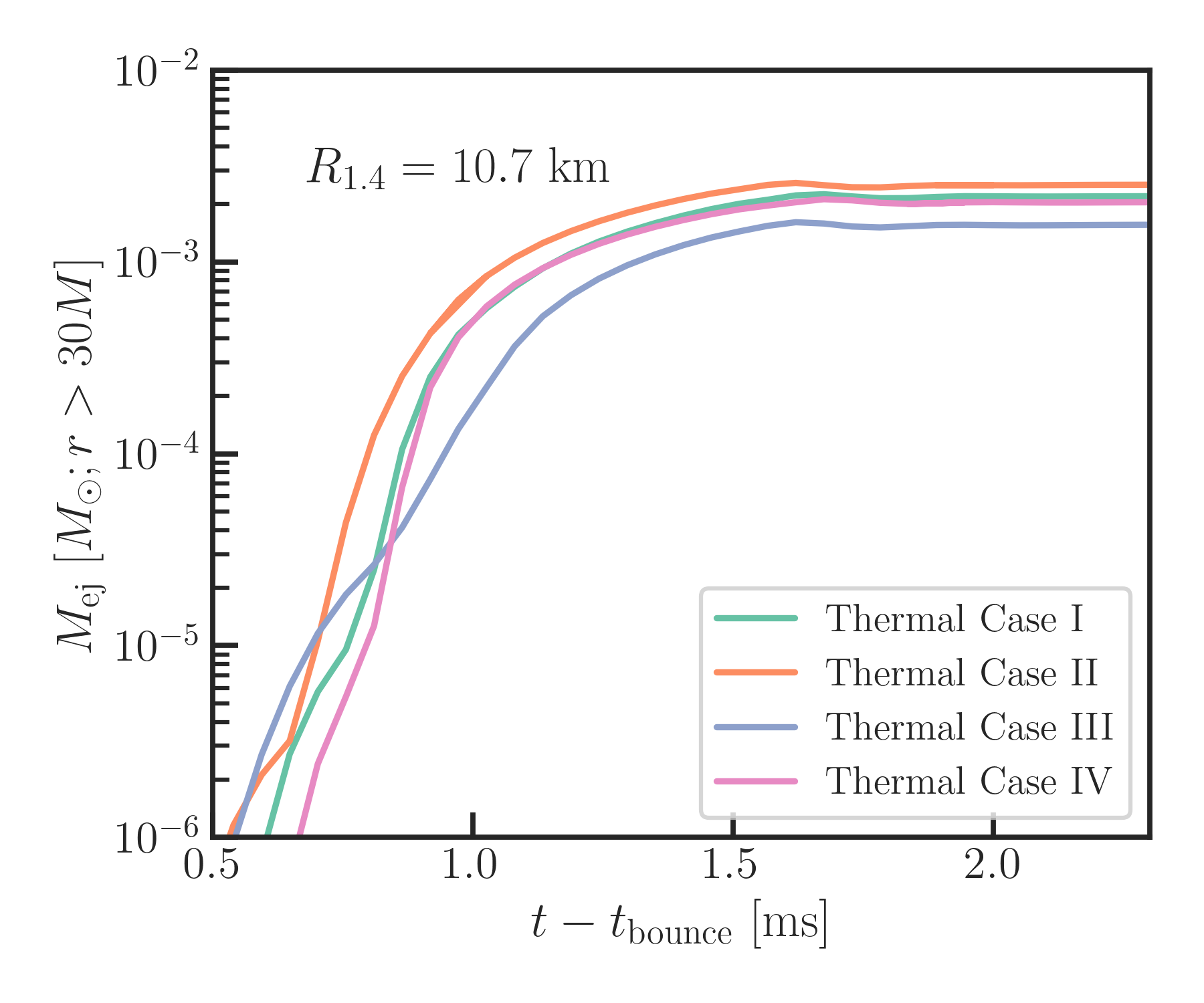} 
\caption{\label{fig:Mej}  Ejecta mass as a function of time since the
start of the bounce, for the bounce-collapse (\Msub) evolutions with 
the \Rsmall cold EoS. }
\end{figure}

We show the evolution of the ejecta mass as a function of time for the
different thermal treatments in Fig.~\ref{fig:Mej}.  At the same late time
(2.3~ms after the start of the bounce), the total amount of
dynamical ejecta ranges from 1.6-2.5$\times 10^{-3}~\Ms$, indicating
44\% fractional differences between the two most different
evolutions (see Table~\ref{table:thermal}).  

For these core-bounce evolutions, we find that the total disk
mass continues to decrease slightly at the end of our simulations,
due to accretion onto the remnant black hole.
To ensure a fair comparison, Table~\ref{table:thermal} 
reports the disk mass
at the same late time ($\sim2.3$~ms after the start of the  bounce)
for all four thermal treatments, and we 
note that the quoted values thus represent an upper limit on 
the total disk mask. We find that the
late-time disk mass ranges from 0.025 to 0.035~$\Ms$ for the
different thermal treatments, indicating 33\% fractional 
difference between the two most different cases. We point out that these differences in $M_{\rm ej}$ and 
$M_{\rm disk}$ are larger than the changes we find as
result of the different Lorentz factor
limits required to evolve some of the bounce-collapse
evolutions through merger (see Apenndix~\ref{sec:caps} for further
discussion).

Determining the numerical significance of these differences would require a resolution study to quantify our errors, which is beyond the scope of the
current work. To provide an approximate reference point,
we note that
in a previous study, we estimated the fractional error in $M_{\rm ej}$
 to be $\sim$30\%,
 based on simulations at three different resolutions,
 for a system with a different cold EoS but a similar total mass of 
 $M_{\rm ADM}= 2.76\Ms$. The resolution of the present
 work is comparable to the medium-resolution from
 that study ($\Delta x=156$), and the initial separations for
 the BNS systems are identical as well  (see \cite{Raithel:2022san} and
 Appendix D of \cite{Raithel:2023zml}).
 Thus, although the conditions of the previous resolution study are
 not an exact match to the current work, we expect $\sim$30\% errors in $M_{\rm ej}$
 to be a reasonable estimate for our current purposes.
 We also point out that these simulations do not include
 neutrinos, which may be important
 for modeling the mass ejection \cite[e.g.,][]{Foucart:2016rxm,Espino:2023mda}. 
 We leave a resolution study under more similar conditions
 and with neutrinos included,
 to investigate these differences in more detail, to a future work.

 \subsection{Results for the stiff cold EoS}
 \label{sec:results_stiff}
 
 We repeat the investigation with a stiffer cold EoS (\Rbig), which
 provides more pressure to support a remnant against collapse, as well
 as more angular momentum at first contact due to the larger NS radii,
 and thus is expected to support a larger threshold mass.  As in
 Sec.~\ref{sec:results_soft}, we simulate a series of binaries of
 varying masses to determine \Mthresh~ via bisection search.  We
 summarize the masses included in the full series, as well as the
 outcome of these evolutions, in Table~\ref{table:masses}.  This
 initial series was evolved with Thermal Treatment I, for which we
 determine the threshold mass for a single core-bounce to be
 \Mthresh$=3.00\Ms$.
   
  We then simulate the binaries with total masses that bound the critical threshold mass
  (i.e., \Msub~ and \Msup; as in Sec.~\ref{sec:results_soft}),
  with the three other thermal treatments and  again
  find no dependence of the threshold mass on the $M^*$-parameters.
  For all cases, we find \Mthresh=3.00$\Ms$.
  
 \begin{figure}[ht]
\centering
\includegraphics[width=0.45\textwidth]{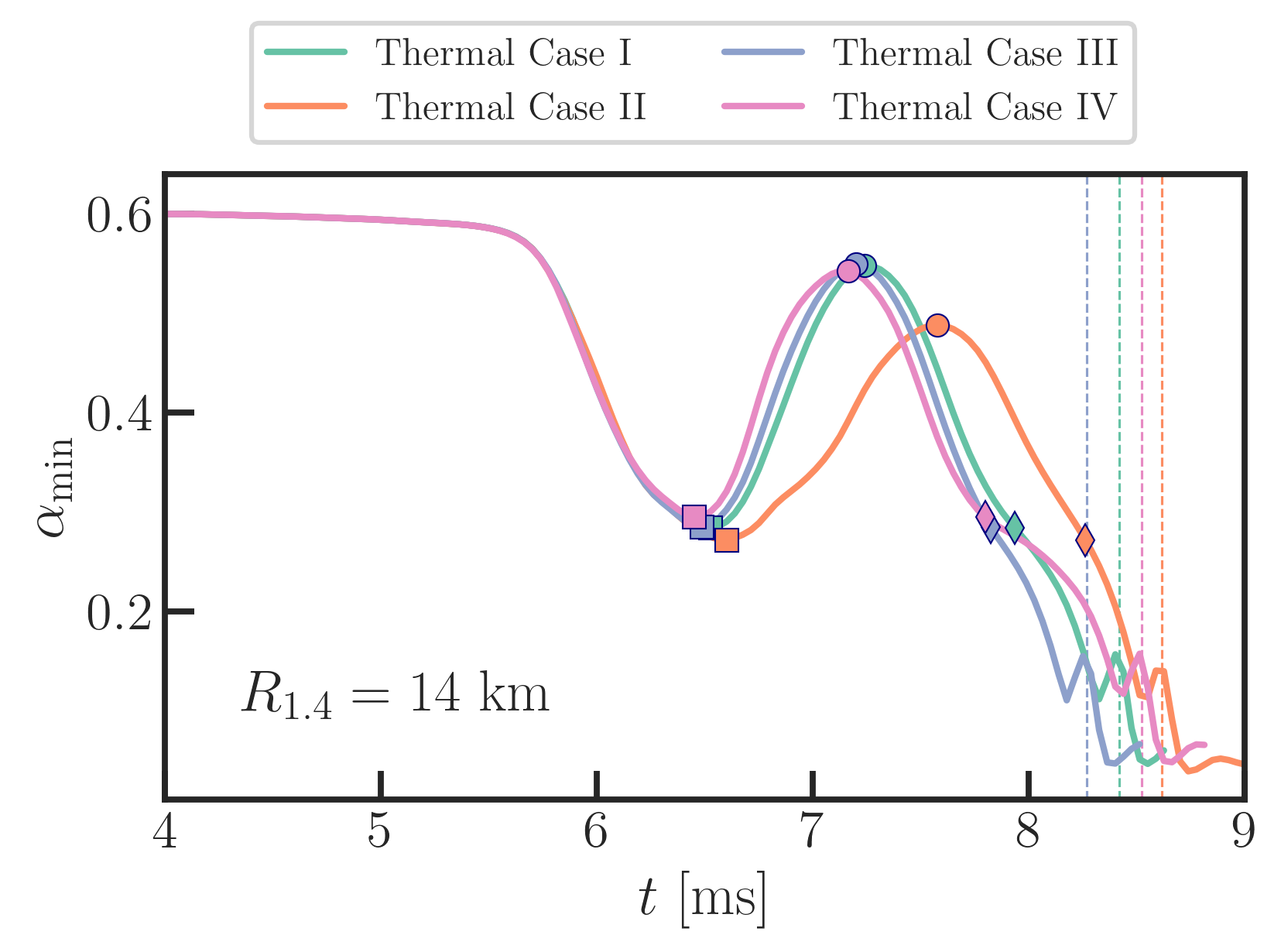}\\
\includegraphics[width=0.45\textwidth]{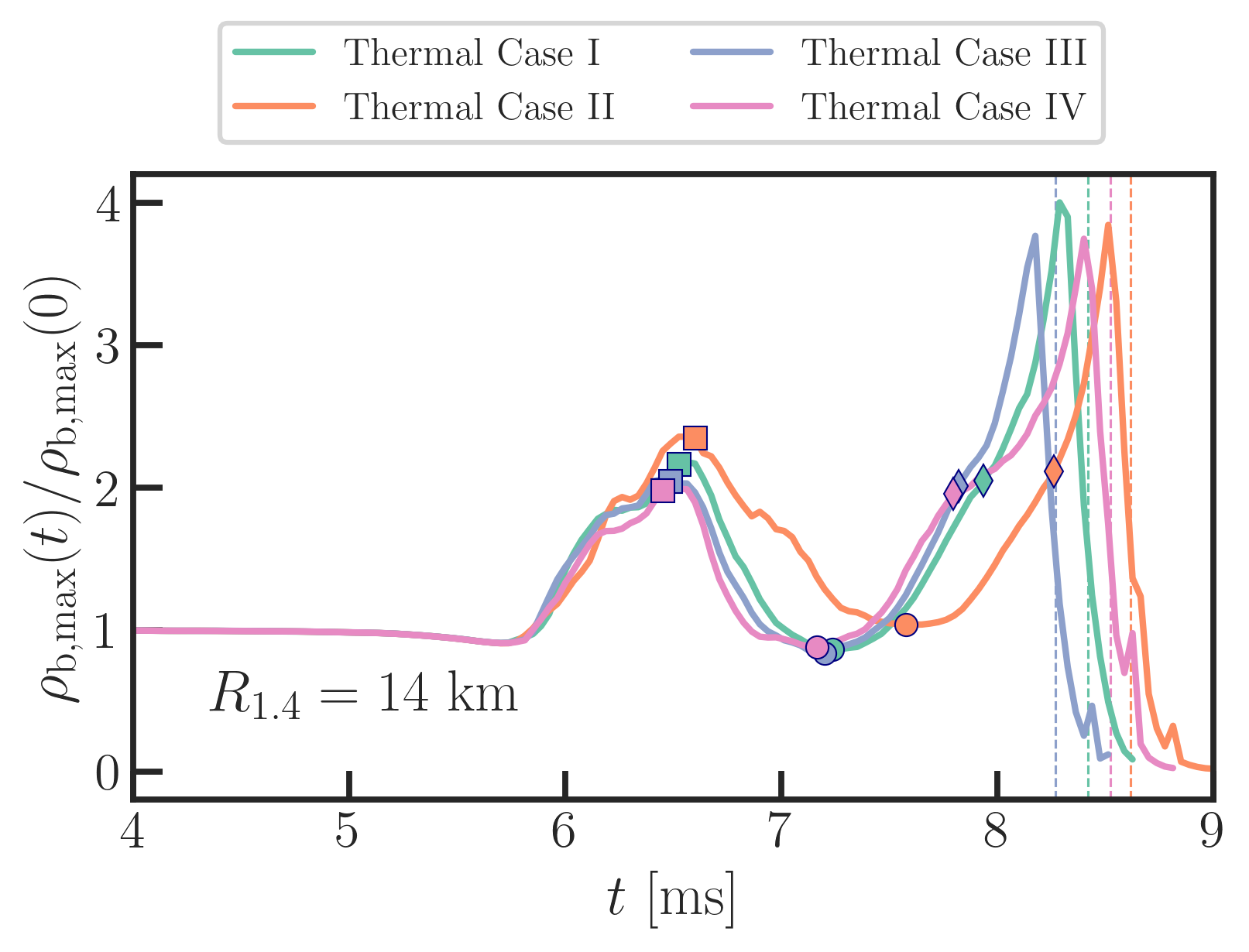}
\caption{\label{fig:lapse_Mthr_R14}  Same as Fig.~\ref{fig:lapse_Mthr_R10}, for the
bounce-collapse (\Msub=2.99$\Ms$) evolutions with the stiff (\Rbig) cold EoS.}
\end{figure}

 \begin{figure}[ht]
\centering
\includegraphics[width=0.45\textwidth]{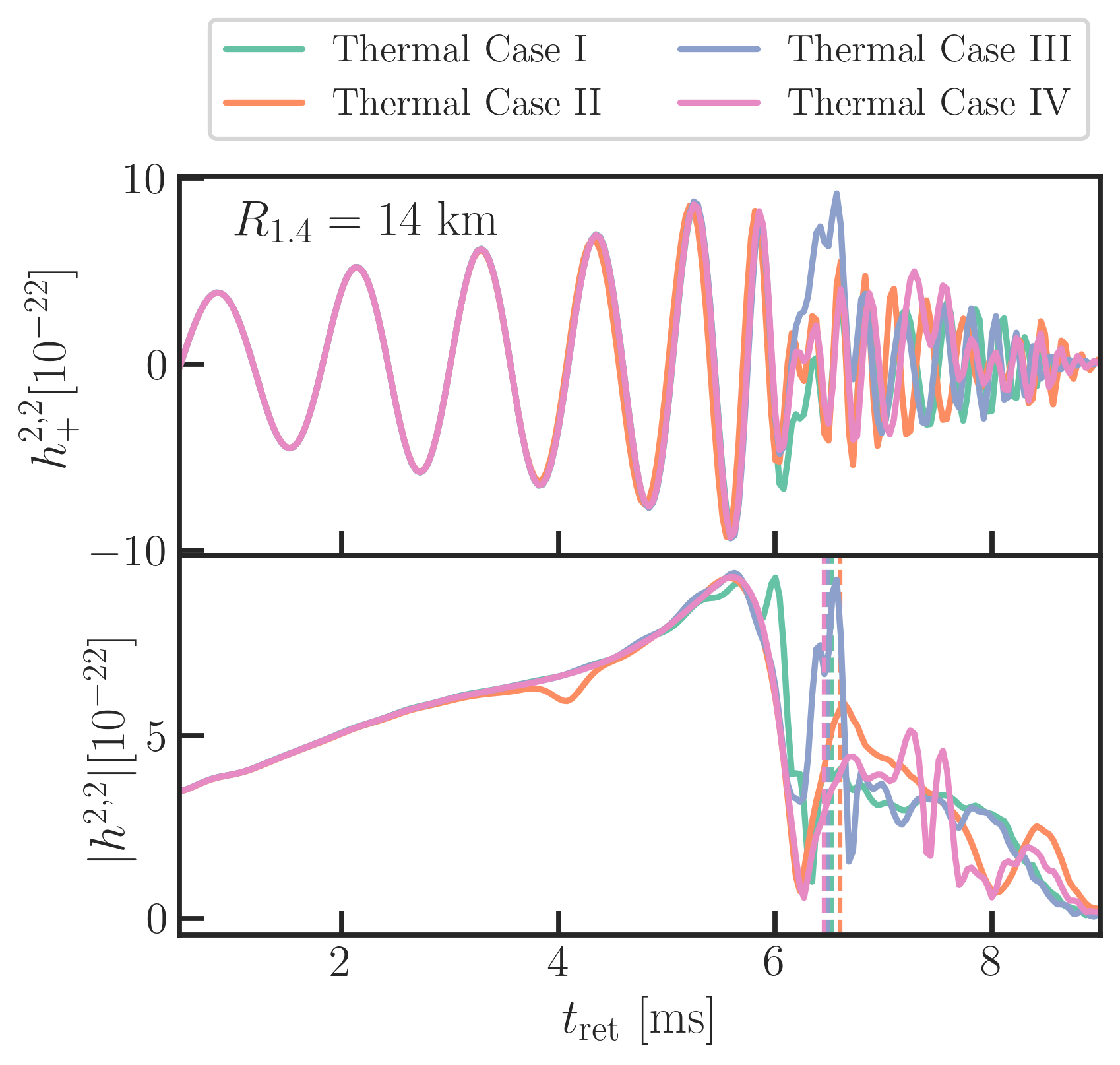}
\caption{\label{fig:gw_R14}  Same as Fig.~\ref{fig:gw_R10}, for the
bounce-collapse (\Msub=2.99$\Ms$) evolutions with the stiff (\Rbig) cold EoS.}
\end{figure}

  \begin{figure*}[ht]
\centering
\includegraphics[width=0.65\textwidth]{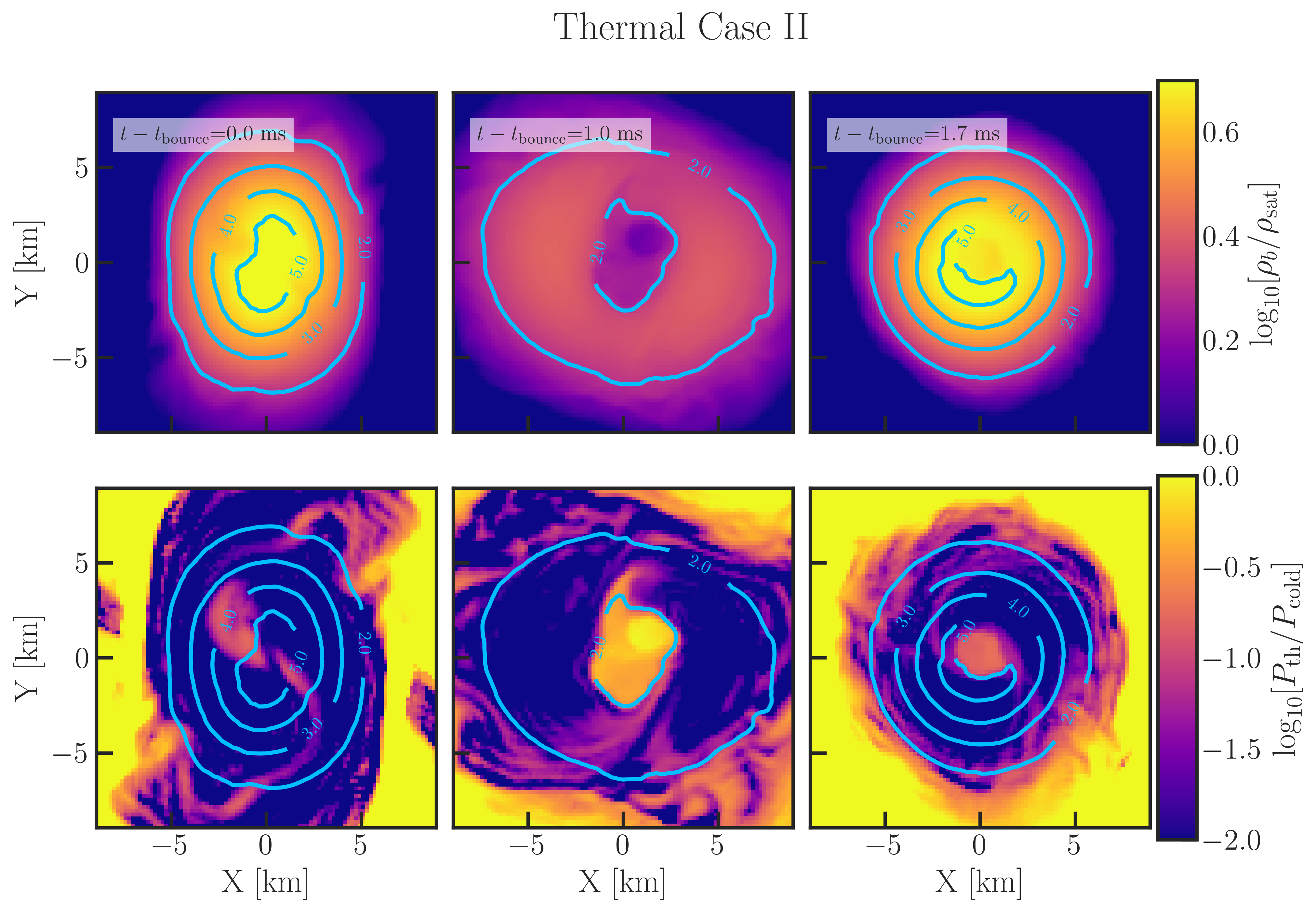}
\caption{\label{fig:Pth_R14_n08a13} Same as Fig.~\ref{fig:Pth_R11_n08a13}, but 
for the \Rbig cold EoS with Thermal Case II.}
\end{figure*}

 \begin{figure*}[ht]
\centering
\includegraphics[width=0.65\textwidth]{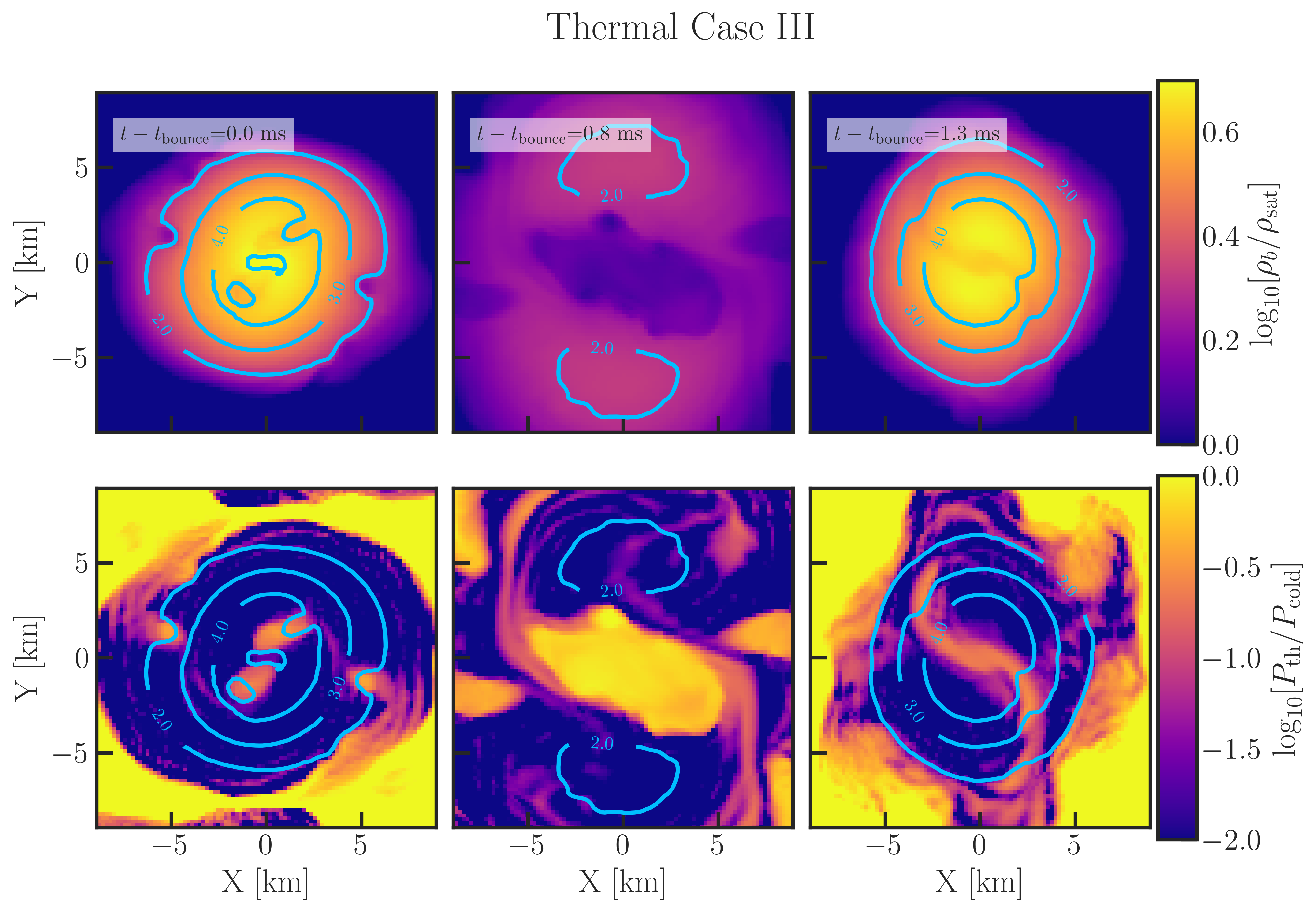}
\caption{\label{fig:Pth_R14_n22a06}  Same as Fig.~\ref{fig:Pth_R11_n08a13}, but 
for the \Rbig cold EoS with Thermal Case III.}
\end{figure*}

\begin{figure*}[ht]
\centering
\includegraphics[width=0.9\textwidth]{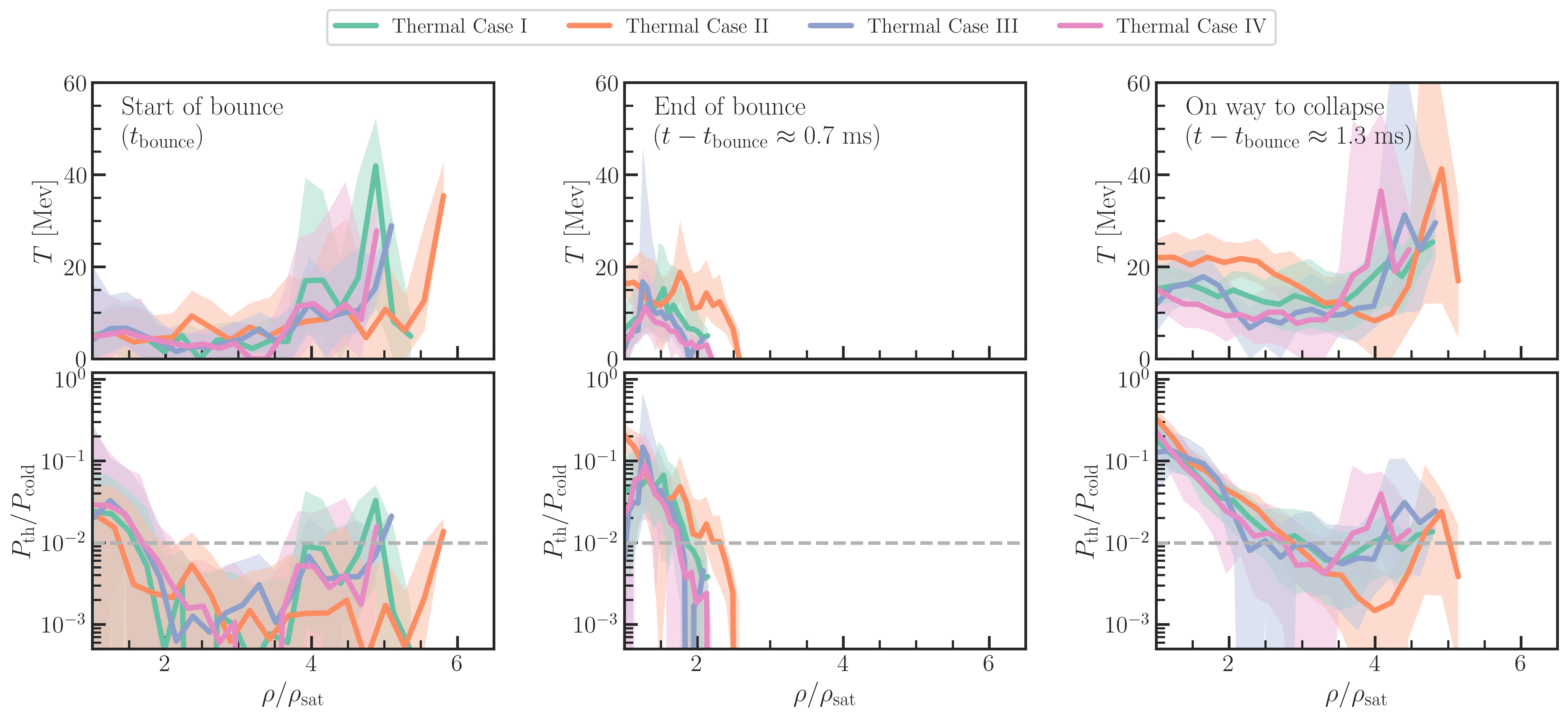}
\caption{\label{fig:medianPth_stiff}  Same as Fig.~\ref{fig:medianPth}, for the
bounce-collapse (\Msub) evolutions with the stiff (\Rbig) cold EoS.}
\end{figure*}

The top panel of Fig.~\ref{fig:lapse_Mthr_R14} shows the minimum of the lapse function
for the bounce-collapse evolutions (\Msub=2.99$\Ms$),
with the \Rbig cold EoS.  As in Fig.~\ref{fig:lapse_Mthr_R10}, 
the markers indicate the
times of the first minimum in \amin (i.e., the start of the bounce),
the end of the bounce, and  one time during collapse.
The bottom panel of Fig.~\ref{fig:lapse_Mthr_R14} shows the compression ratio for the same evolutions. As was the case for the soft cold EoS, the lapse-defined times bracketing the bounce correspond to local maxima and minima in the compression ratio, as expected. Interestingly, we find that three of the four thermal cases with the \Rbig cold EoS lead to a $\sim2\times$ compression of the maximum rest-mass density, at the start of the bounce. This is the same compression ratio at the bounce onset that we found for the bounce-collapse evolutions with the \Rsmall cold EoS, despite the different softness/stiffness of the EoSs. This hints at a universality in the dynamics of the bounce-collapse evolutions, across at least these two very different EoSs.

We  show the GWs for these \Msub~ evolutions in Fig.~\ref{fig:gw_R14}. The time of the  of the bounce onset is shown with the dashed vertical dotted line. As was the case for the soft EoS, the lapse-definition of $t_{\rm bounce}$ is approximately associated with a local peak in the GWs. This can be explained by the fact that the remnant is most compact at the bounce onset, and then the cores move apart making the remnant less compact.
  
From the minimum lapse function, we find that Thermal Case II leads to slightly 
weaker and longer core bounce than the other three thermal treatments. 
In particular, the bounce  ends $\sim$0.4~ms later
and the value of \amin~ at the peak corresponding to end of the bounce is reduced by $\sim$11\% as well,
hinting at a weaker bounce overall. 
This can also be seen in the compression ratio of the rest-mass density, where the matter reaches a larger initial compression ratio of 2.4, and remains
more compressed to later times.

  \subsubsection{Thermal profiles for the bounce-collapse evolutions with the stiff cold EoS} 
  \label{sec:thermal_stiff}
To further investigate these differences,
Figs.~\ref{fig:Pth_R14_n08a13} and \ref{fig:Pth_R14_n22a06}
show snapshots of the density and thermal pressure profiles
at the start of the bounce, the end of the bounce,
and on the way to collapse, for the Thermal Cases II and III, respectively. 
In the middle column, Fig.~\ref{fig:Pth_R14_n08a13}
for Thermal Case II clearly shows a suppressed bounce,
most visible in the blue contours of constant rest-mass density,
which do not form separate lobes, as seen in 
Fig.~\ref{fig:Pth_R14_n22a06} with Case III
(and with both other thermal treatments as well).
  
Figure~\ref{fig:medianPth_stiff} 
shows the median thermal properties for the bounce-collapse evolutions, for 
each thermal treatment.  As in Fig.~\ref{fig:medianPth}, 
the median quantities are calculated within uniformly-spaced
density bins, extracted from 2D snapshots at the 
start of the bounce, the end of the bounce, and
during the collapse.  Figure~\ref{fig:medianPth_stiff} 
illustrates that  the Case II thermal treatment
(shown in orange) experiences slightly less heating 
at the bounce onset, as indicated by a 
lower median thermal pressure at high densities. This reduced early heating may explain
  the weaker bounce observed from the lapse in
  Fig.~\ref{fig:lapse_Mthr_R14} and from
  the spatial snapshots in Fig.~\ref{fig:Pth_R14_n08a13}. 
 The weaker bounce also leads to higher
 core densities in the middle column of 
  Fig.~\ref{fig:medianPth_stiff} (i.e., by the end of the bounce).
In particular, by the end of the bounce, we find that the average
density of the supranuclear core is 1.82$\rns$ for Thermal Case II,
while the cores evolved with the other thermal treatments have
an average density of 1.53-1.56$\rns$ at the same time 
(see Table~\ref{table:thermal}).  
  
  We thus find that, although
  thermal effects are not strong enough to change the
  threshold mass, the differences in the thermal
  evolution can still have a small impact on the strength
  and duration of the core bounce for certain
  thermal treatments, with the stiff cold EoS.
   
 \subsubsection{Dynamical ejecta for the stiff cold EoS} 
  \label{sec:ejecta_stiff}
\begin{figure}[ht]
\centering
\includegraphics[width=0.45\textwidth]{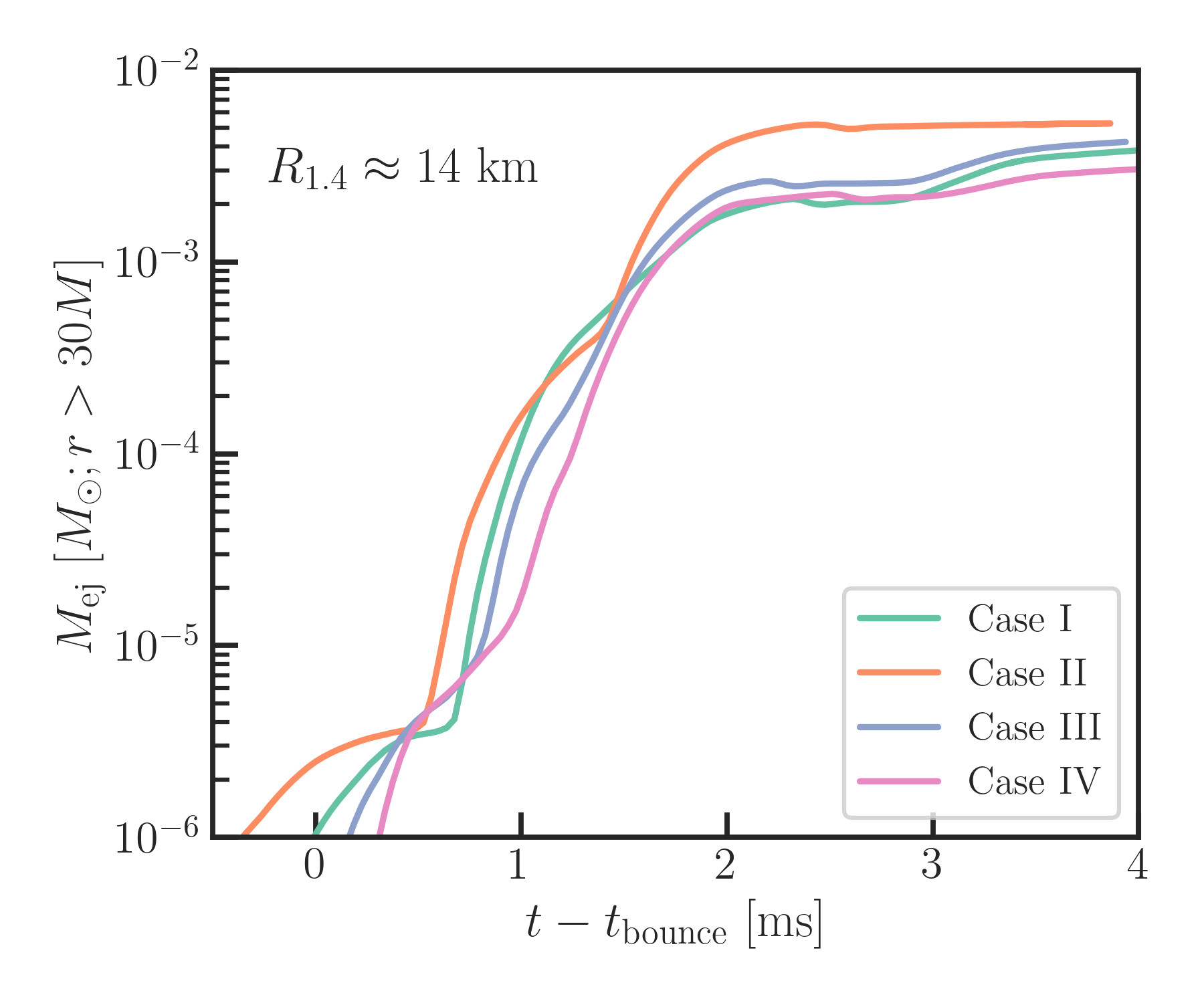} 
\caption{\label{fig:Mej_stiff}  Same as Fig.~\ref{fig:Mej}, for the
bounce-collapse (\Msub) evolutions with the stiff (\Rbig) cold EoS.}
\end{figure}

  In terms of observable signatures,
the weaker core bounce is associated with enhanced 
 dynamical ejecta. 
 We show the dynamical ejecta for all bounce-collapse
 evolutions for the \Rbig cold EoS, with the four different thermal
 treatments, in  Fig.~\ref{fig:Mej_stiff}.
We summarize the total $M_{\rm ej}$ and upper limits
on the disk mass $M_{\rm disk}$ for each case in 
Table~\ref{table:thermal}. For a fair comparison,
we calculate and report the ejecta and disk mass at 
the same relative coordinate time of 3.9~ms after the bounce.

The Case II evolution, which leads to the weaker
core bounce observed above, results in up to $\sim60\%$
more dynamical ejecta than the other thermal treatments.
Although this thermal case required a more aggressive Lorentz factor
limit to evolve through merger, we find that this
cannot fully explain the enhanced ejecta for this case.
Indeed, in our tests, we find that changing the
Lorentz factor limit for this cold EoS leads to only 20\%
fractional differences in $M_{\rm ej}$ 
(see Appendix~\ref{sec:caps}). 
  
 By contrast, we find only a modestly enhanced disk mass for this case. Overall,
 we find disk masses of $\sim 0.032-0.038~\Ms$,
 indicating up to 17\% differences in $M_{\rm disk}$
 as a result of the thermal treatment.

\section{Discussion and conclusions}
\label{sec:discussion}
Quasi-universal relations between the threshold mass for prompt
collapse and properties of the cold EoS-- such as \Mtov~ and the
stellar radius -- as well as predictions for the disk mass left
outside the remnant BH
formed following a BNS merger at the threshold mass 
provide a promising way of constraining the EoS at
high densities.  However, any such constraints are limited by how
accurately empirical relations can capture the true breadth of
outcomes.

In this work, we have systematically investigated one source of
uncertainty for BNS mergers near the threshold for prompt collapse, due
to imperfect knowledge of the finite-temperature part of the EoS. We
find that the threshold mass for prompt collapse is insensitive to the
details of the (density-dependent) thermal prescription, to within
0.625\% in \Mthresh, for the two cold EoSs and four thermal treatments
studied here.

We note that $0.625\%$ is the error in the iterative procedure for determining the threshold mass. Determining whether this is the absolute error, within the physics assumptions adopted in our work, would require a resolution study, which goes beyond the scope of the current study, but will be included in a forthcoming follow-up work.
We note that
previous work by Ref.~\cite{Koppel:2019pys}, using codes similar to ours,
performed simulations with the same EoS at multiple resolutions and obtained threshold masses differing only within a variance of $\sim0.005 M_{\odot}$. Thus, we expect that the main conclusions and trends found in our work -- i.e., that the threshold mass for prompt collapse is not sensitive finite temperature effects -- are robust with resolution.

Beyond the numerical resolution, we also
do not expect other code choices such as our Riemann solver, reconstruction scheme, or atmospheric treatment to significantly affect the collapse thresholds  reported here. The impact of these choices was first studied in the context of prompt collapse threhsolds in Ref.~\cite{Baiotti:2010ka}, in which the authors compared the \texttt{Whisky} and \texttt{SACRA} codes, which differ in both the local Riemann solver and the atmosphere treatment. Despite those differences, for configurations chosen that promptly collapse, they report black holes masses that differ by at most 0.15\%, and black hole dimensionless spins that differ by at most 1\% (or even agree to better than 0.1\% in one case) \cite{Baiotti:2010ka}.

We also presented a first look at the thermal features of sub-threshold mergers, which undergo a single core-bounce during their collapse.
During a bounce-collapse, we find that the merging core can become
significantly heated, with median temperatures up to 10-20~MeV by the
end of the bounce, depending on the thermal treatment.  These
temperatures correspond to median thermal pressures of $\sim 1-5$\% of
the cold pressure at 2$\rns$, for these EoSs. This highlights that
there is modest heating that happens during the bounce, but that the
degree of heating is generally not very sensitive to the details of
the thermal treatment (i.e., to the choice of the $M^*$-parameters in
the framework from \cite{Raithel:2019gws}). The exception is for one of
our cases (Thermal Case II with the \Rbig cold EoS), for which this
choice of $M^*$-parameters leads to a slightly suppressed and longer
core bounce, compared to the other three thermal treatments.  Although
the effect is small, it highlights that the differences in the thermal
profile of the merging core can have an impact on the dynamics of the
bounce-collapse, even if it is not enough to alter the threshold mass.

The bounce-collapse evolutions produce up to $1.6-5.3\times10^{-3}\Ms$ of
dynamical ejecta and disk masses of 2.5-3.8$\times10^{-2}\Ms$, 
depending on the cold EoS and thermal treatment adopted
in the simulations. We find fractional differences of 
$\lesssim40\%$ in the ejecta and disk masses, due to the thermal treatment
for most cases; but up to a $\sim60\%$ fractional increase in the dynamical
ejecta for the \Rbig cold EoS with Thermal Case II, which
experienced the weaker core bounce.

We thus find a correlation between the weaker core
bounce for the stiff cold EoS with Thermal Case II, and an
increase in the dynamical ejecta. This could possibly be
explained by the more compressed remnant that initially forms,
which speeds up as it compresses and then launches 
faster ejecta as it expands. Alternatively, the small
differences in heating with this thermal treatment
may lead to stronger shocks that launch the additional ejecta.
We plan to investigate this further in a future work.

The possible sensitivity of the ejecta and disk masses
to the thermal prescription is of
particular interest, because some lower limits on the neutron star tidal
deformability (or radius) have been inferred from the ejecta of
GW170817. In particular, Ref. \cite{Radice:2017lry} compared
estimates of the disk mass inferred from
kilonova AT2017gfo against a set of BNS merger
simulations at the same total mass. Based on these simulations, the
authors tentatively excluded EoSs with binary tidal deformability
$\widetilde{\Lambda} < 400$, which was later revised to
$\widetilde{\Lambda} < 300$ \cite{Radice:2017lry,Radice:2018ozg}
\cite[but see][who weakened this constraint
significantly]{Kiuchi:2019lls}. The simulations in the 
present work are a first step
towards quantifying the uncertainties in the dynamical ejecta and disk mass onto the remnant black hole due
purely to uncertainties in the thermal physics, for masses near the
threshold mass. These findings motivate studies of thermal effects in asymmetric bounce-collapse BNS mergers, which can have stronger shock heating and larger disk masses~\cite{Kiuchi:2019lls}, as different thermal treatments could lead to even larger disks and hence lower the constraint on the minimum binary tidal deformability of GW170817, and the constraints it implies on the nuclear EoS. This will be the topic of future work of ours.

We have attempted to bracket a wide range of cold EoSs with the choice
of the \Rsmall and \Rbig models. However, we note that it remains
possible that some exotic EoSs may be more susceptible to differences
in the thermal treatment, for binaries near the threshold
mass. Further work would be needed to extend our results, e.g. to EoSs
with strong phase transitions.  Additionally, further work would be
needed to investigate whether our results would change for binaries with non-zero initial spins.
We note also that the present simulations
do not include neutrinos, which may be important for accurately
capturing the mass ejection
\cite[e.g.,][]{Foucart:2016rxm,Espino:2023mda}. 
Nevertheless,
this study provides one step forward towards quantifying uncertainties
due to the thermal physics in determining threshold mass for prompt
collapse, and it provides a first look into the thermal evolution of
bounce-collapse mergers, for binaries near the threshold mass evolved
with realistic thermal prescriptions.

\begin{acknowledgments}
This work was supported, in part, by NSF Grant PHY-2145421 and NASA
Grant 80NSSC24K0771 to the University of Arizona.  The simulations
presented in this work were carried out with the Anvil cluster at the
Purdue University Rosen Center for Advanced Computing and the
Stampede3 cluster at the Texas Advanced Computing Center, under XSEDE
allocation PHY190020.  The simulations were also performed, in part,
with the Firebird Computing Cluster, supported by Swarthmore College
and Lafayette College.
\end{acknowledgments}

\bibliography{ms,non_inspire}
\bibliographystyle{apsrev4-1}

\appendix

\section{Generalized piecewise polytropic fit for the \Rsmall cold EoS}
\label{sec:appendixEOS}

  \begin{table*}
  \caption{\label{table:GPP} Best-fit parameters for the generalized piecewise polytropic
representation of the soft cold EoS. 
$R_{1.4}$ indicates the radius of a 1.4~$\Ms$ neutron star predicted by each EoS,
$\Lambda_{1.4}$ gives the tidal deformability at the same mass, and
the remaining columns
 provide the eight free parameters that are determined via our GPP
 fitting procedure. }
  \centering
\begin{tabular}{ccccccccccc }
\hline 
EoS  &  $R_{1.4}$ [km]  & $\Lambda_{1.4}$ &  $\log_{10} K_1$  &   $\Gamma_1$   &  $\Gamma_2$ &  $\Gamma_3$ &  $\Gamma_4$&  $\Gamma_5$  &  $\Gamma_6$  &  $\Gamma_7$\\
\hline  
Soft & 10.6 & 214 & -49.162 & 4.224 &  0.0879 &  14.032 &  4.133 &  0.984 &  2.319 &   2.426 \\
\hline
\end{tabular}
 \end{table*}

In this appendix, we describe the construction of the \Rsmall cold EoS.
The motivation for this cold EoS is to approximately match
the softest model from Ref.~\cite{Kiuchi:2019lls}
 that produced enough ejecta in numerical simulations to reproduce
 the observed luminosity of AT2017gfo \cite{LIGOScientific:2017ync}.
 
 In that work, the softest consistent model was constructed from
 three piecewise-polytropic segments \cite{Read:2008iy}.
 The low-density segment is defined by a polytropic index $\Gamma=1.357$ and
 polytropic coefficient $K=3.584\times10^{13}$ in cgs units.
 The middle-density segment is described by $\Gamma=4.007$,
 with the pressure at a density of $10^{14.7}$ g/cm$^3$
 given by $\log_{10}P$ (dyn cm$^{-2}$) = 34.1. The middle
 segment is matched to the low-density EoS at the density
 where the pressures intersect, which for this model
 is at 2.982$\times10^{14}$g/cm$^3$. Finally, the high-density polytropic
 segment is specified by $\Gamma=2.8$ at densities above 
 10$^{15}$ g/cm$^3$ \cite{Kiuchi:2019lls}.

 We fit this three-polytrope model with the generalized piecewise polytrope
 (GPP) parametrization of  \cite{OBoyle:2020qvf}. The details of
 the fit are similar to that described in Appendix A of \cite{Raithel:2022san}
 and \cite{Raithel:2023zml};  we recreate the key details below.
 
 In the GPP parametrization, the pressure along a segment is given by
 \begin{equation}
P(\rho) = K_i \rho^{\Gamma_i} + \Lambda_i, \quad	\rho_{i-1} < \rho \le \rho_i,
\end{equation}
where the coefficient $K_i$ is determined by requiring differentiability,
\begin{equation}
\label{eq:Ki}
K_i = K_{i-1} \left( \frac{\Gamma_{i-1}}{\Gamma_i} \right) \rho_{i-1}^{\Gamma_{i-1}-\Gamma_i},
\end{equation}
and the parameter $\Lambda_i$ is imposed to ensure continuity, according to
\begin{equation}
\label{eq:Lambdai}
\Lambda_i = \Lambda_{i-1} + \left( 1- \frac{\Gamma_{i-1}}{\Gamma_i} \right) K_{i-1} \rho_{i-1}^{\Gamma_{i-1}} .
\end{equation}

In our GPP parametrization, we place fiducial densities $\rho_i$ 
at each of the original dividing densities 
($\rho=2.982 \times10^{14}$g/cm$^3$ and  10$^{15}$ g/cm$^3$).
To improve the accuracy of the fit, we place additional fiducial densities
10\% above and below each of these values, for a total of six fiducial densities, 
or seven total segments.

For this seven-segement parametrization, there are 8 free parameters:
$K_1$ and $\Gamma_{1-7}$. All other $K_i$ and $\Lambda_i$
in Eqs.~\ref{eq:Ki}-\ref{eq:Lambdai} are uniquely determined from these values.

To determine the values of these parameters that minimize
the least-squares difference between the three-polytrope \Rsmall
model and its GPP representation, we perform a Markov Chain 
Monte Carlo simulation. We report the resulting best-fit coefficients in
Table~\ref{table:GPP}.

The best-fit parameters for the stiff (\Rbig) cold EoS are provided in Appendix A of
\cite{Raithel:2023zml}.

\section{Impact of the maximum Lorentz factor}
\label{sec:caps}

In order to ensure that velocities do not become unphysical, especially in cases where matter collapses to form a black hole,
numerical relativity codes typically limit the velocity
with a maximum Lorentz factor, $\gamma$, for the fluid,
as measured by a normal observer \cite[e.g.,][]{Chang:2020ktl}.
In our code, we typically set this limit to
$\gamma_{\rm max}=50-100$.  However, in the \Msub~
evolutions with some thermal treatments, allowing
the velocity to increase to very large values leads 
to code failures and the evolutions cannot
proceed past black hole formation.  
For cases where we encounter the code failures, we instead implement 
$\gamma_{\rm max}= 15$. This more aggressive limit
on the Lorentz factor ensures numerical stability and
successfully allows the
evolutions to proceed past the collapse to form a black hole.

\begin{figure}[ht]
\centering
\includegraphics[width=0.4\textwidth]{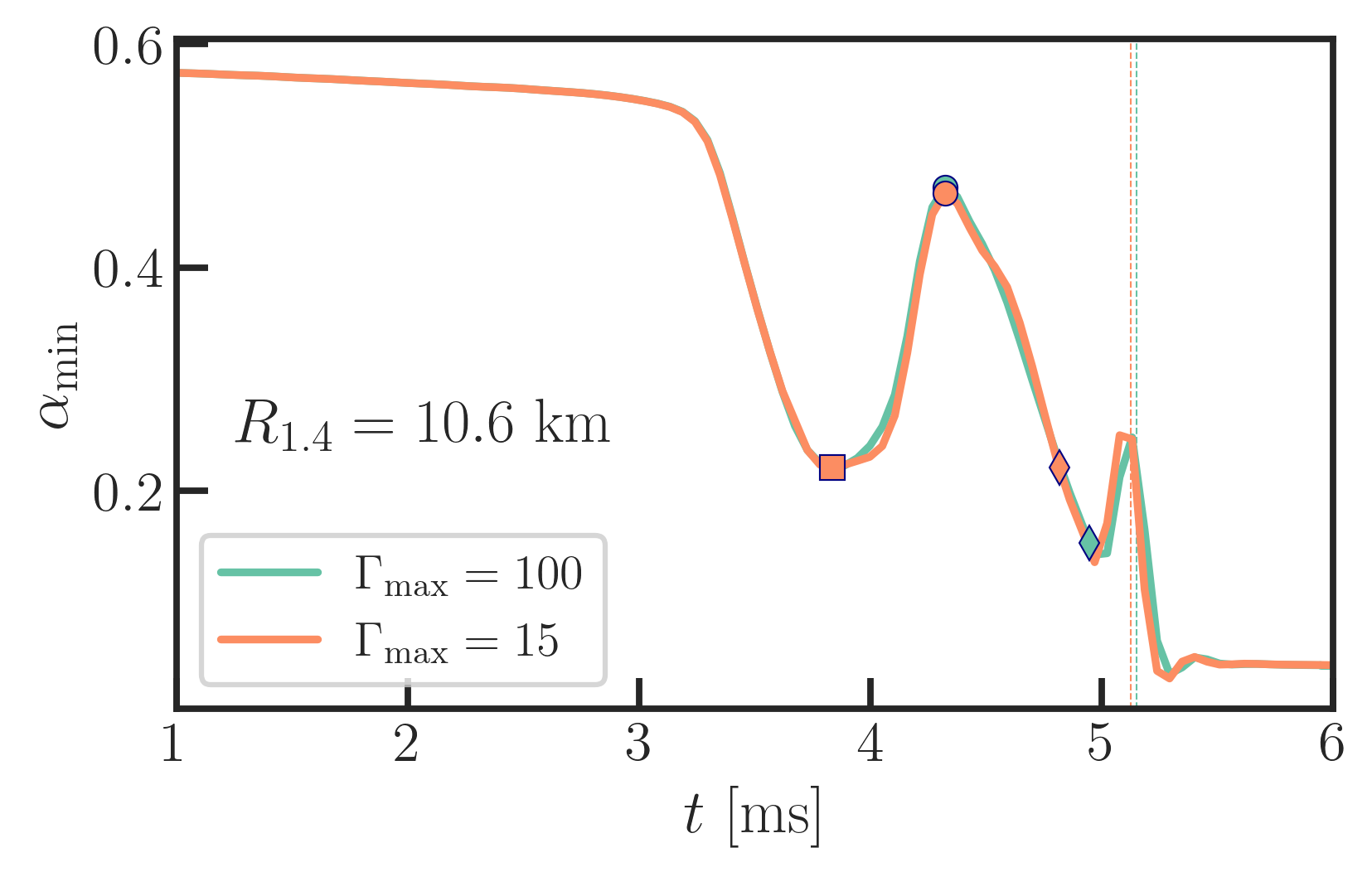}
\includegraphics[width=0.4\textwidth]{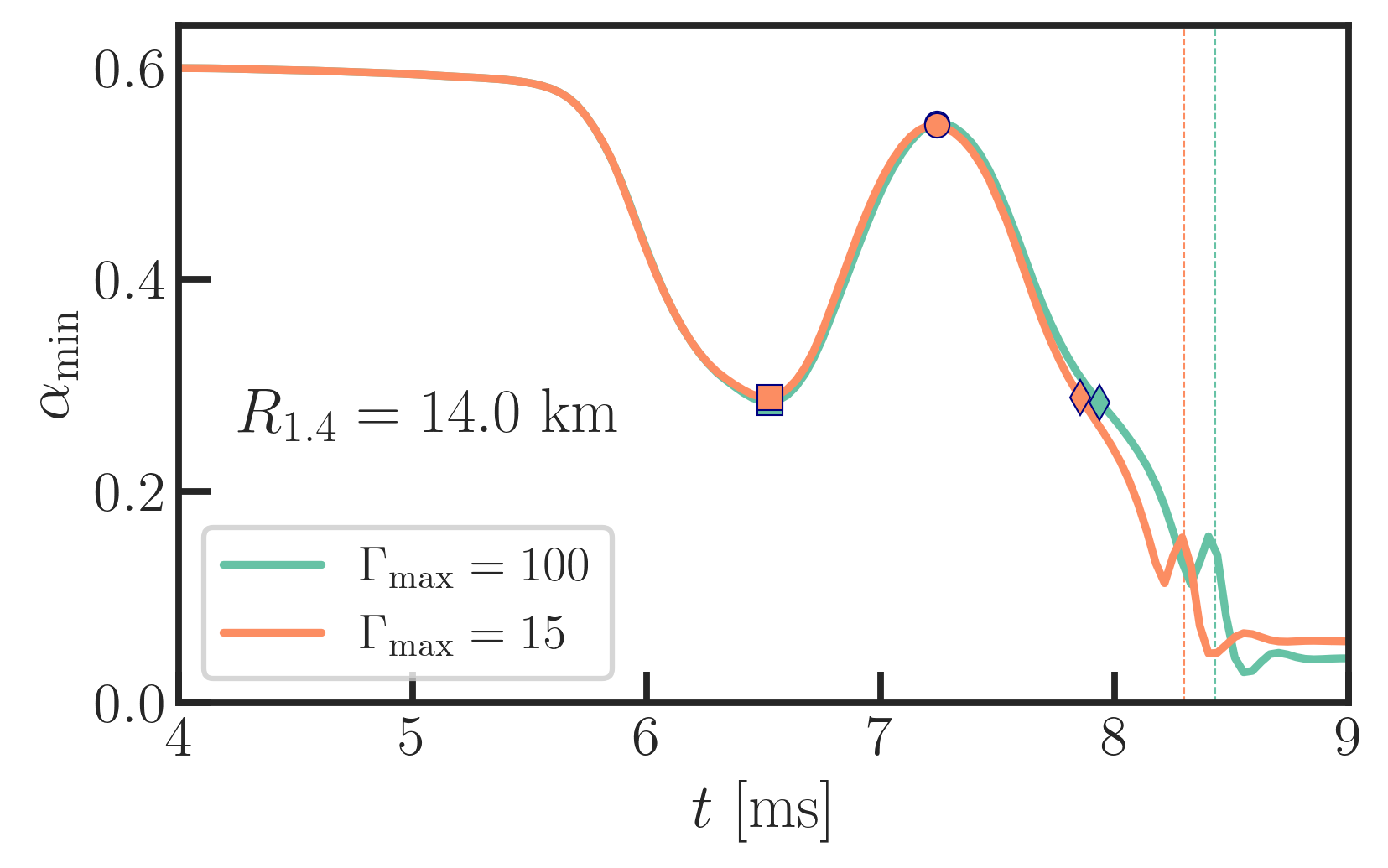}
\caption{\label{fig:lapse_Pcap}  Minimum lapse for the bounce-collapse
 evolutions with the \Rsmall cold EoS (top) and \Rbig cold EoS (bottom), 
evolved with Thermal Case I and the two different caps on the Lorentz factor.
The standard cap ($\gamma_{\rm max}=100$) is shown in teal, while the
more aggressive cap (at $\gamma_{\rm max}=15$) is shown in orange.
The vertical dotted line indicates the time at which the apparent horizon
is first found.}
\end{figure}

\begin{figure}[ht]
\centering
\includegraphics[width=0.4\textwidth]{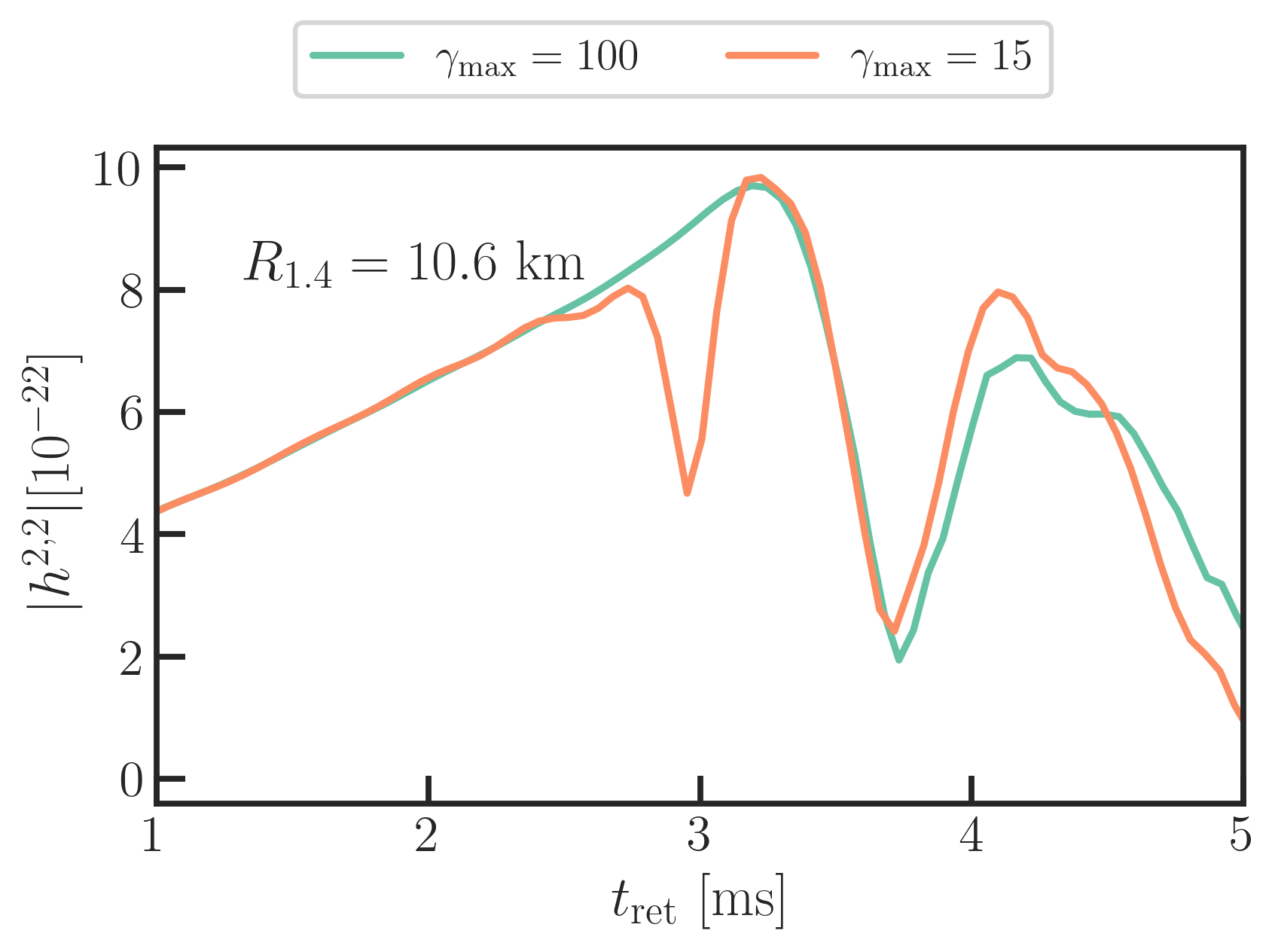}
\includegraphics[width=0.4\textwidth]{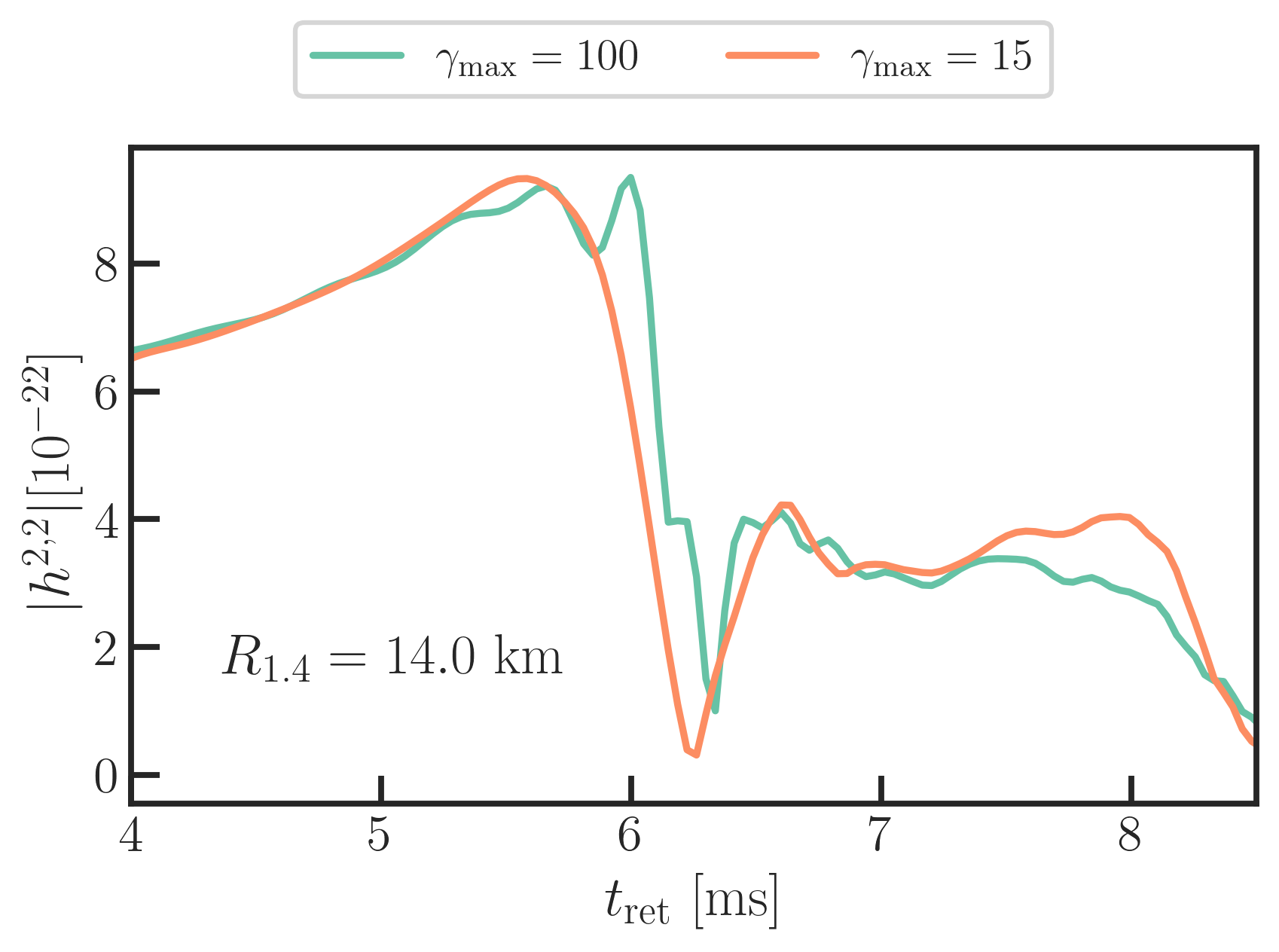}
\caption{\label{fig:gw_Pcap} Amplitude of the $\ell=m=2$ mode of the
gravitational wave strain for the bounce-collapse
 evolutions with the \Rsmall cold EoS (top) and \Rbig cold EoS (bottom), 
evolved with Thermal Case I and the two different Lorentz factor caps.
The standard cap ($\gamma_{\rm max}=100$) 
is shown in teal, while the
more aggressive cap ($\gamma_{\rm max}=15$) is shown in orange. }
\end{figure}

\begin{figure}[ht]
\centering
\includegraphics[width=0.4\textwidth]{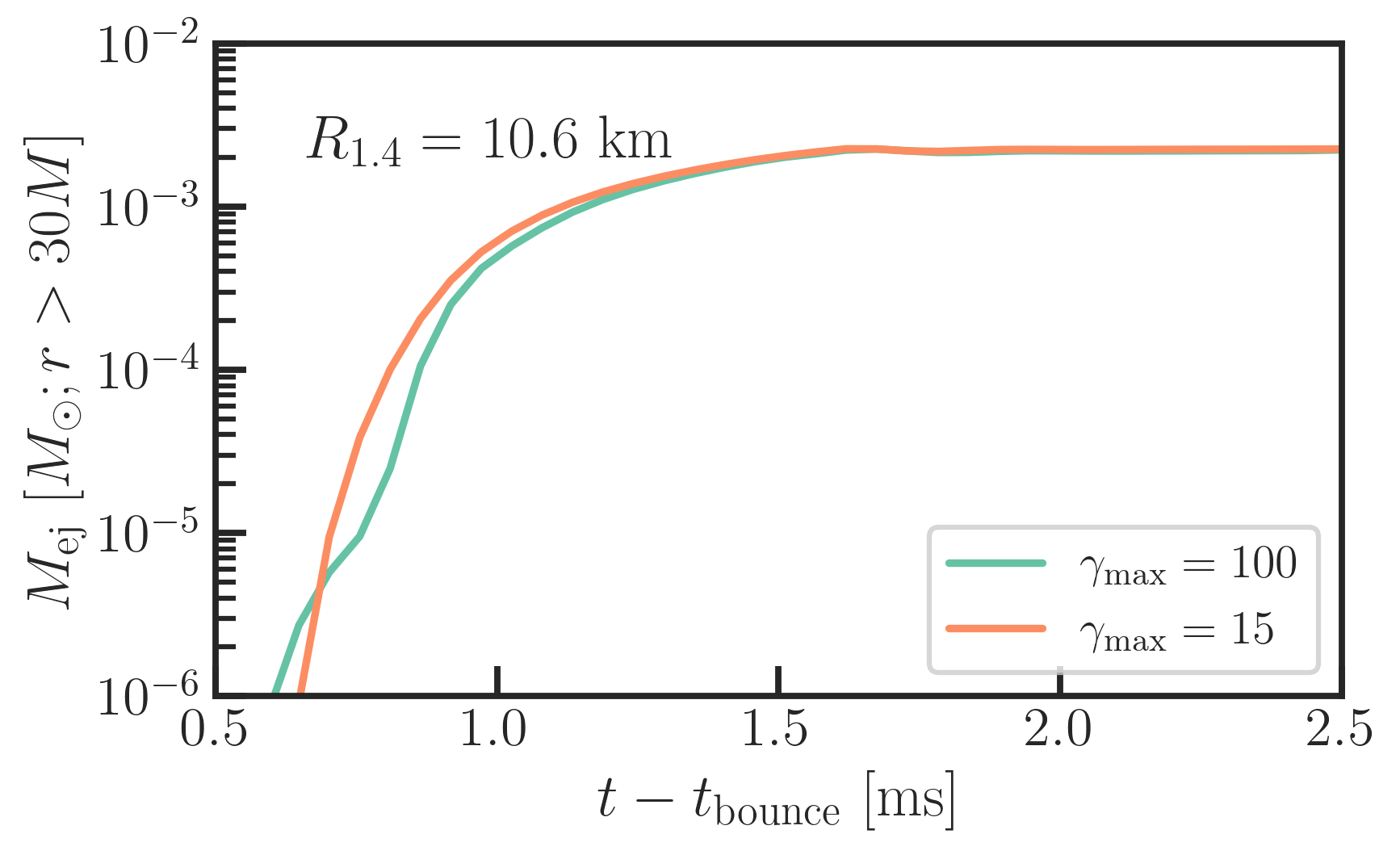}
\includegraphics[width=0.4\textwidth]{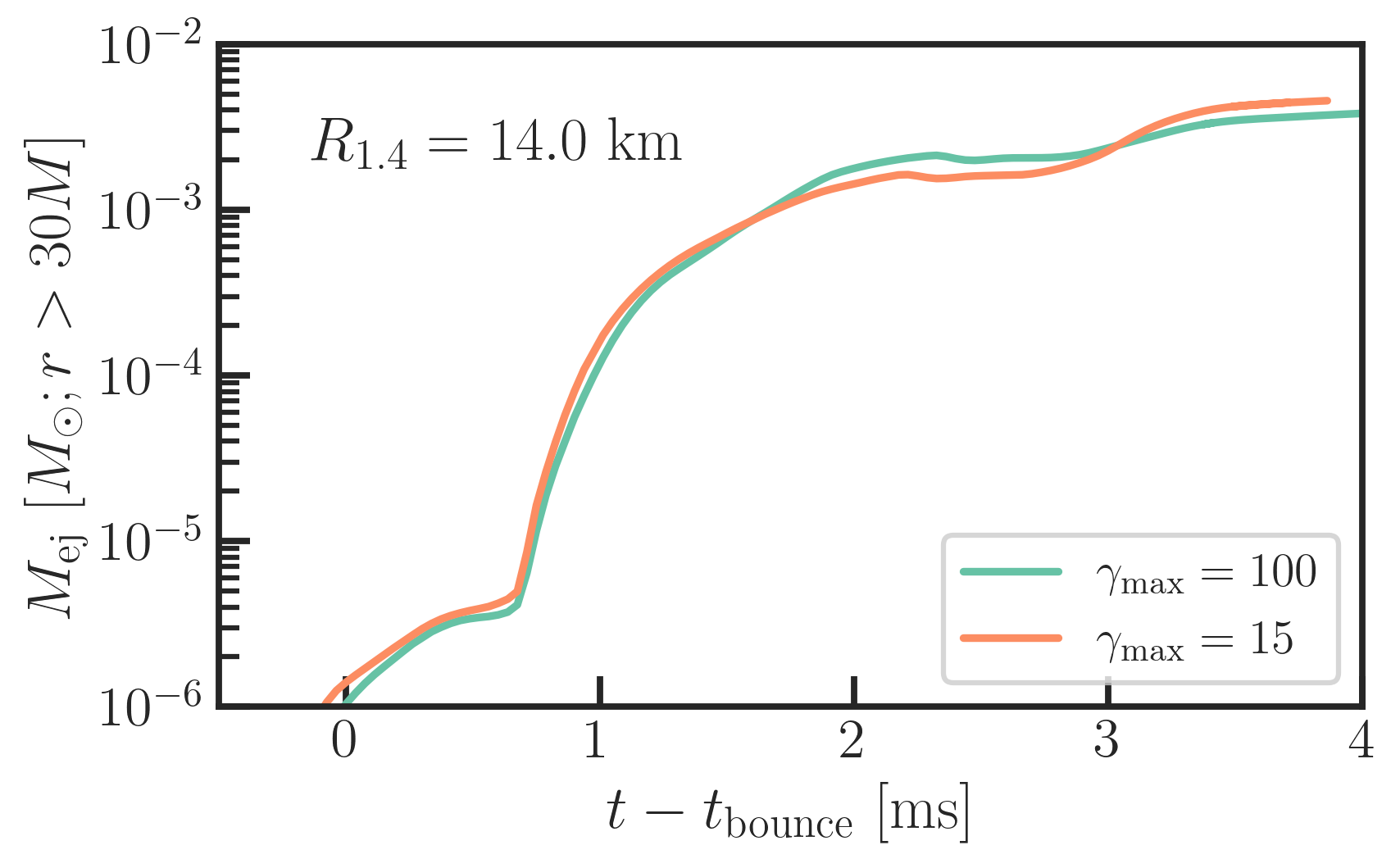}
\caption{\label{fig:ej_maxgam} Dynamical ejecta mass as a function
of time since the start of the bounce, for the bounce-collapse evolutions
with the \Rsmall cold EoS (top) and \Rbig cold EoS (bottom),
with both Lorentz factor caps. We find negligible differences in
the dynamical ejecta as a result of the Lorentz factor cap. }
\end{figure}

In this work, we use $\gamma_{\rm max}=100$, unless we encounter
issues with the numerical stability, in which case we adopt
the more aggressive limit of $\gamma_{\rm max}=15$.
We use the more aggressive limit for the \Msub~ evolutions for
the \Rsmall cold EoS with Thermal Treatments II and IV, and for the
\Rbig cold EoS with Thermal Treatment II. 

To confirm that reducing the Lorentz factor limit does not affect the results
presented in this paper, we performed a comparison of Lorentz factor caps
for a subset of \Msub~ simulations that we are able to perform with
either choice of cap. In particular, we simulated the \Msub~ evolution
with the \Rsmall and \Rbig cold EoSs, with Thermal Case I (which is
well-behaved in all cases), and both the standard and aggressive caps.

 We show the comparison of the minimum lapse function, \amin,
 with the two Lorentz factor caps in
 Fig.~\ref{fig:lapse_Pcap}. For both cold EoSs, we find
 negligible differences in \amin~ as a result of the Lorentz factor limit.
 In particular, we find no difference in the time
 of the bounce for both EoSs and $\lesssim1\%$
 difference in the peak amplitude of \amin~ during the core bounce
for either cold EoS. 
 
 The corresponding GW strain amplitude is shown in Fig.~\ref{fig:gw_Pcap}, which shows differences of $<1\%$ in the maximum GW amplitude for either cold EoS.
 
 These differences 
 are considerably smaller than the 0.4~ms delay in the end of the bounce 
 and 12\% reduction of \amin~ during the peak bounce
 that was found in the main paper for the
 \Rbig cold EoS with Thermal Case II. 
 In other words, the suppressed core
 bounce found in the main paper for the \Rbig
 evolution with Thermal Case II cannot
 be explained by the choice of $\gamma_{\rm max}$.

We show the dynamical ejecta for these tests
in Fig.~\ref{fig:ej_maxgam}. At a time of $\sim2.3$~ms after the bounce, we find a fractional difference of $2\%$ in the
dynamical (unbound) ejecta for the \Rsmall cold EoS, and $23\%$ in the (bound) disk mass. For the \Rbig cold EoS at a
time $3.9$~ms after the bounce, there is a 20\% fractional difference in dynamical ejecta and 2\% in the disk mass. These differences are smaller than maximum differences we find as a result of the thermal treatment in the main paper, 
for each cold EoS.

\begin{figure*}[ht]
\centering
\includegraphics[width=0.9\textwidth]{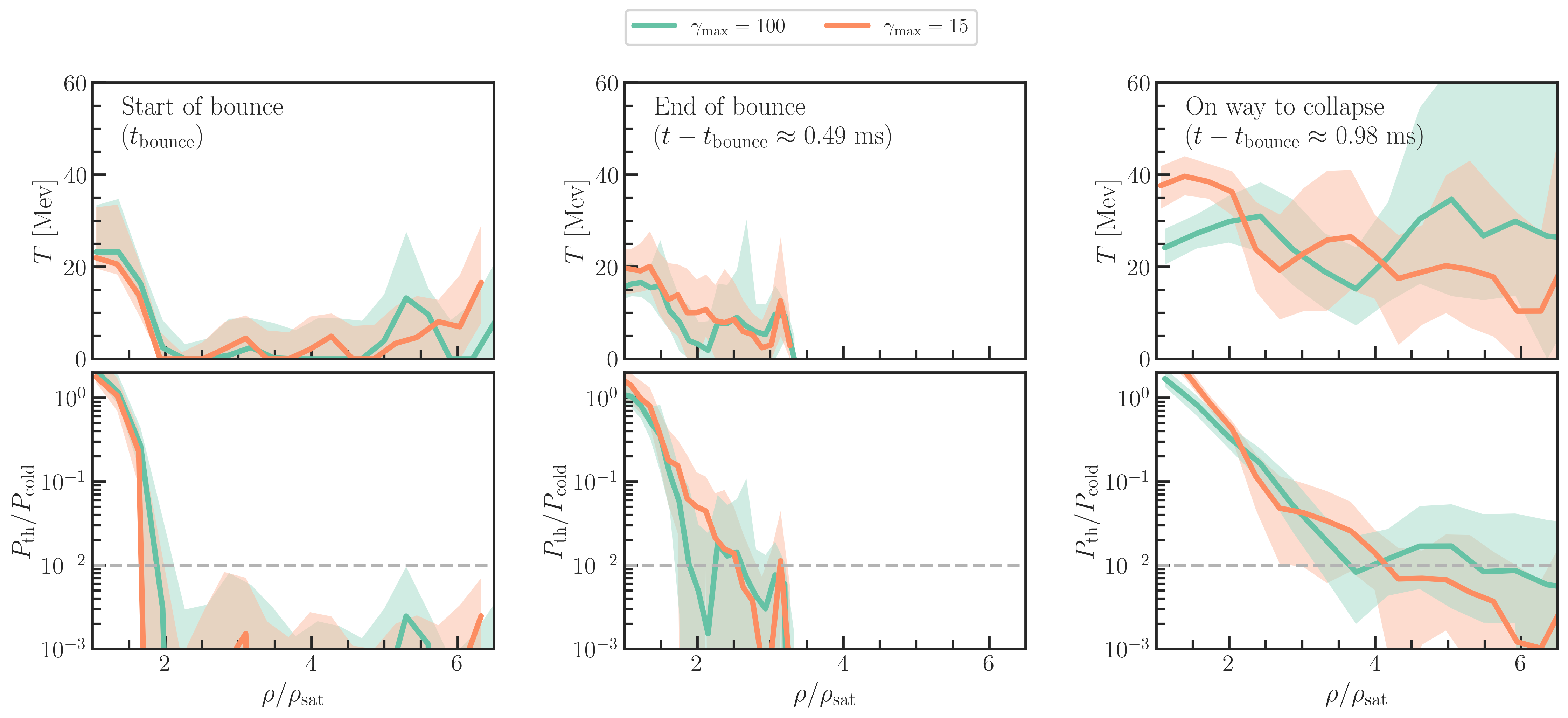}
\caption{\label{fig:medianPth_caps_R10} Same as Figure~\ref{fig:medianPth}, but for the 
\Rsmall cold EoS with Thermal Treatment I and comparing the two Lorentz factor caps. 
As in Fig.~\ref{fig:medianPth}, the median thermal quantities are
calculated from 2D snapshots at three different times, using uniformly-spaced density bins. 
The shaded regions indicating 68\% bounds.}
\end{figure*}

\begin{figure*}[ht]
\centering
\includegraphics[width=0.9\textwidth]{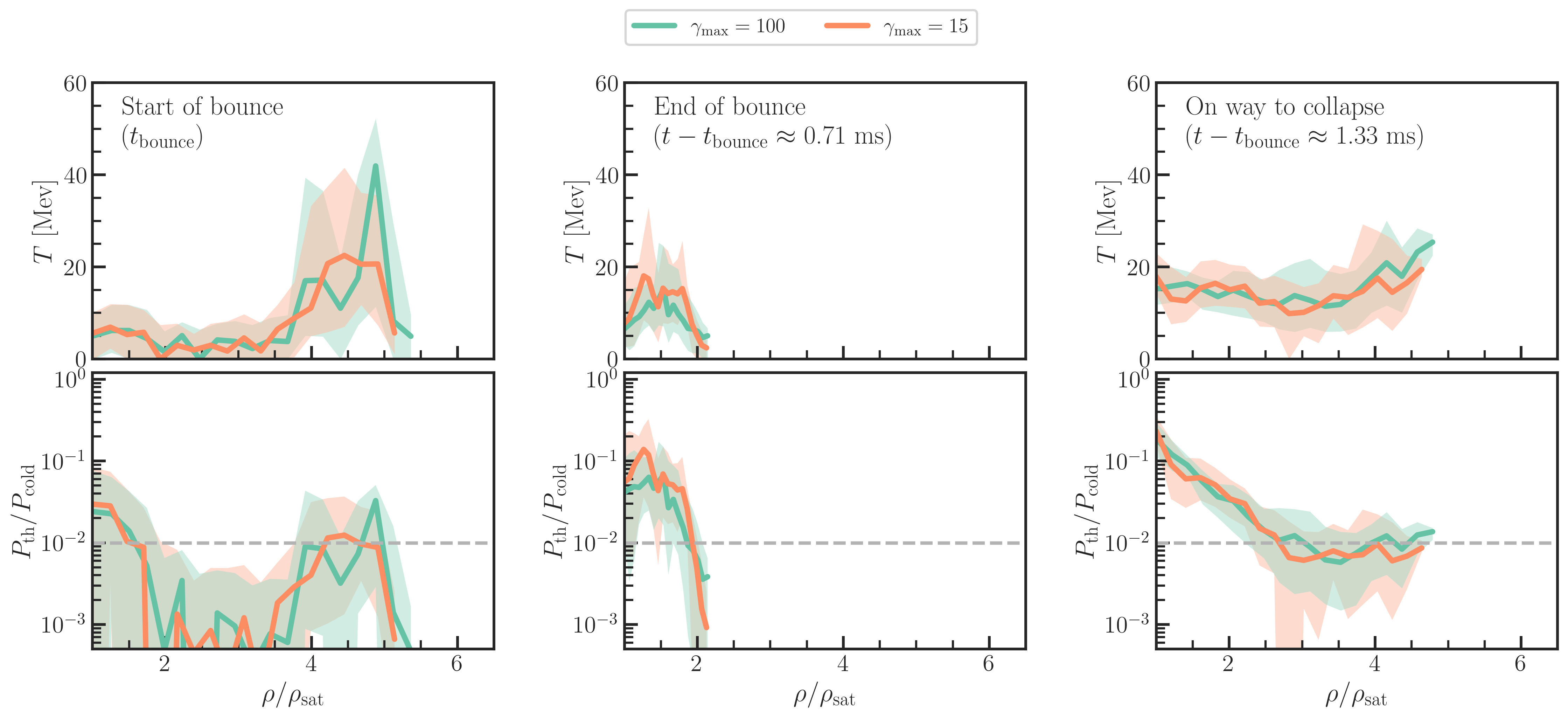} 
\caption{\label{fig:medianPth_caps_R14} Same as Figure~\ref{fig:medianPth_caps_R10}, but for the 
\Rbig cold EoS.}
\end{figure*}

 To explore how the Lorentz factor limit affects the thermal properties,
 Figs. \ref{fig:medianPth_caps_R10}-\ref{fig:medianPth_caps_R14}
  shows the median thermal quantities at the time
  of the start of the bounce, the end of the bounce, and one time during the
  collapse (indicated by markers in Fig.~\ref{fig:lapse_Pcap}),
 for the evolutions with Thermal Case I and each Lorentz factor cap.
There are negligible differences in the thermal pressure
and temperature of the supra-nuclear matter at the start of the bounce 
and at the late-time snapshot. At the end of the bounce, the evolution
with the reduced Lorentz factor limit has a slightly higher median
temperature and thermal pressure profile. However, the
$1-\sigma$ regions overlap for both Lorentz factor caps,
and the shape of the thermal profiles are similar.
Thus, we conclude that the Lorentz factor
limit has a negligible impact on the thermal profiles
for the supra-nuclear matter of the merging cores.

In summary, our evolutions with Thermal Case I and
the two Lorentz factor limits indicate that the
outcome of the evolution (i.e., the bounce-collapse)
is not sensitive to the Lorentz factor limit. Additionally,
there are negligible differences in the dynamics
and thermal evolution of the dense-matter core,
and relatively small differences in the
ejecta and disk masses, between
the two Lorentz factor caps used. In other words,
these tests confirm that the results of
the main paper are not significantly affected
by changes to the Lorentz factor limit.

\section{Estimation of composition effects}
\label{sec:composition}
In this work, we have neglected changes to the composition,
instead assuming that the matter retains its initial 
electron fraction, $Y_{e,~\rm{cold}~\beta}(n)$, 
as determined by the conditions of cold, neutrinoless 
$\beta$-equilibrium \cite{Raithel:2019gws}.
Because the timescales of a prompt collapse are too fast 
for neutrinos to cool the matter, it is safe to ignore neutrinos
(and thus compositional changes) from that point of view. 
However, out-of-equilibrium effects may nevertheless 
be important in the determination of the threshold
for prompt collapse, as deviations from the initial 
$Y_{e,~\rm{cold}~\beta}(n)$ can change the pressure profile
of the merging system. In this appendix, we estimate
the magnitude of composition effects for the bounce-collapse
simulations of this work. We find that out-of-equilibrium
effects contribute $\lesssim1-2\%$ fractional
change to the total pressure, compared to the thermal-only
treatment that we adopt in this work. We
conclude that focusing on thermal effects is thus a reasonable
first step towards quantifying the role of 
second-order EoS effects on the prompt collapse threshold
(i.e., going beyond the cold EoS),
though a full treatment including neutrino physics will be
necessary to provide a final answer.

To start, we note that we can generically
break the total pressure into a cold,
$\beta$-equilibrium contribution,
plus some correction that brings the matter to 
finite-temperature and arbitrary composition, 
according to
\begin{equation}
\label{eq:pressure}
P(n,T,Y_e) = P(n,T=0, Y_{e,~\rm{cold}~\beta}) + P_{\rm corr}.
\end{equation} 

 In the $M^*$-framework used in this work, we
neglect any compositional changes and assume 
the matter always has the same $Y_e(n)$, 
as determined by the initial conditions of 
cold, neutrinoless $\beta$-equilibrium. 
This amounts to a correction term of
\begin{equation}
\label{eq:dP_our}
	P_{\rm corr; ~\rm our ~model}=  
        P(n,T, Y_{e,~\rm{cold}~\beta}) –
        P(n,T=0, Y_{e,~\rm{cold}~\beta}), 
\end{equation}
which approximates the full (temperature- and composition-dependent)
correction with a thermal-only model.

For estimating the impact of composition changes,
we assume that neutrinos
do not have time to re-equilibrate the matter for
mergers near the threshold for prompt collapse. The assumption is motivated by the fact that neutrinos are trapped and do not have enough time to escape within the collapse timescale in our remnants.
In this case, the electron fraction is advected with the matter
and thus can deviate from its initial value. 
We take an extreme limit to further bound the 
problem: that the electron fraction is everywhere
increased by a constant amount. We choose increases
of 0.01 or 0.02, above the initial $Y_{e,~\rm{cold}~\beta}$,
as motivated by $Y_e$ profiles
from simulations that include neutrinos and
evolve with similar ranges of EoSs as explored in this
work \cite[e.g.,][]{Most:2021ktk,Radice:2023zlw}. In this case,
the complete (temperature- and composition-dependent)
pressure correction in Eq.~(\ref{eq:pressure}) is
\begin{equation}
\label{eq:dP_adv}
P_{\rm corr;~advection} = P(n,T, Y_{e,~\rm{cold}~\beta} + \xi) – P(n,T=0, Y_{e,~\rm{cold}~\beta})
\end{equation}
where $\xi=(0.01, 0.02)$.

\begin{figure}[ht]
\centering
\includegraphics[width=0.45\textwidth]{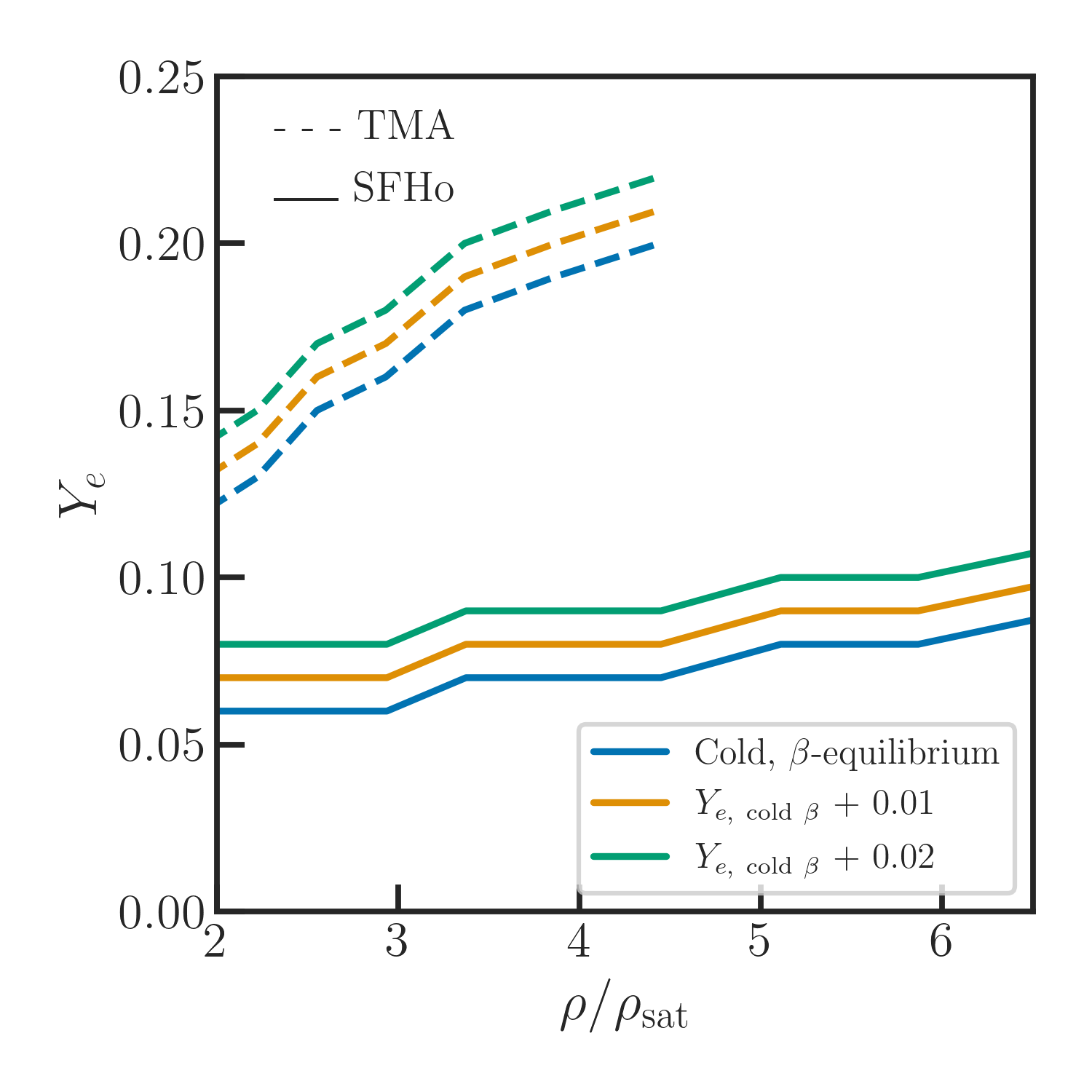} 
\caption{\label{fig:Ye} Electron fraction
profiles explored in this appendix, for the 
SFHo (solid) and TMA (dashed) tabulated EoSs. 
The elctron fraction for cold, $\beta$-equilibrium 
matter is shown in blue,
and constant increases of 0.01 and 0.02 above this are
shown in orange and green, respectively.}
\end{figure}

Our goal is to quantify how much our thermal-only model 
is missing, compared to the “full” correction for 
these two example cases where the composition also changes.
To address this, we use the tabulated 3D EoSs SFHo
\cite{Steiner:2012rk} and TMA \cite{Toki:1995ya}
as a representative soft and stiff EoS,
to stand in for the parametric models used in our work.
Using the 3D tables allows us to test the 
importance of compositional effects independently from the approximations of the $M^*$-framework. These models have $R_{1.4}$=11.9 and 13.8 km, respectively, 
and symmetry energy slopes of $L$=47 and 90 MeV, very close to the $M^*$-framework EoSs we adopt.

We take the median temperature profile for
our bounce-collapse simulations from the
late-time snapshots in Figs.~\ref{fig:medianPth} and 
\ref{fig:medianPth_stiff},
for the evolutions with Thermal Treatment I. 
This snapshot corresponds roughly
to the time when the stellar cores have merged 
after the bounce, and reached their
initial maximum rest-mass density from the bounce
for the second time (see Figs.~\ref{fig:lapse_Mthr_R10}
and \ref{fig:lapse_Mthr_R14}).
For these snapshots, we show the three different
$Y_e$ profiles under consideration in Fig.~\ref{fig:Ye}.
We show matter from 2$\rns$ to the core
density of each snapshot, in order
to focus on the high-density matter most relevant for
countering the collapse.

For the temperature-density profiles from
these snapshots and each of the possible
$Y_e$ profiles in Fig.~\ref{fig:Ye}, we
calculate the pressure corrections 
using either the SFHo or TMA EoS tables,
to represent the \Rsmall and \Rbig simulations,
respectively.

\begin{figure}[ht]
\centering
\includegraphics[width=0.45\textwidth]{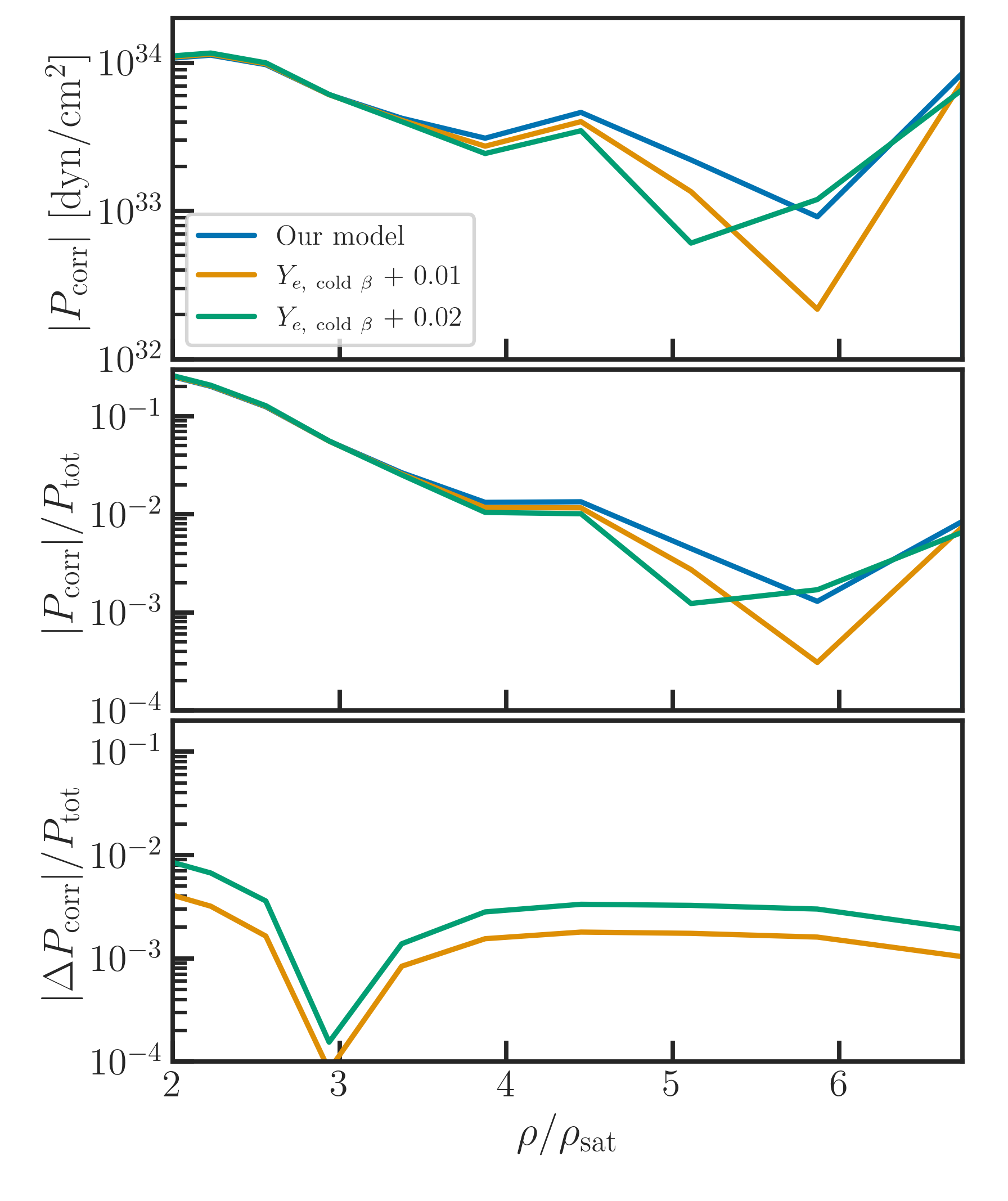} 
\caption{\label{fig:Pcorr_soft}  Pressure corrections
to the cold, $\beta$-equilibrium EoS for the
late-time median temperature profile for the \Rsmall
bounce-collapse simulation. SFHo is used as a 
realistic 3D EoS table, to calculate the full
temperature- and composition-dependent pressures.
Our thermal-only
model used in this work is shown in blue; constant
increases to the initial electron fraction
are shown in orange and green. Top: the absolute
change to the cold, $\beta$-equilibrium pressure
for these thermal-only and thermal-and-composition
corrections. Middle: Fractional changes,
relative to the total pressure. Bottom: Fractional 
differences between the thermal-only model and
the thermal-and-composition dependent models.} 
\end{figure}

\begin{figure}[ht]
\centering
\includegraphics[width=0.45\textwidth]{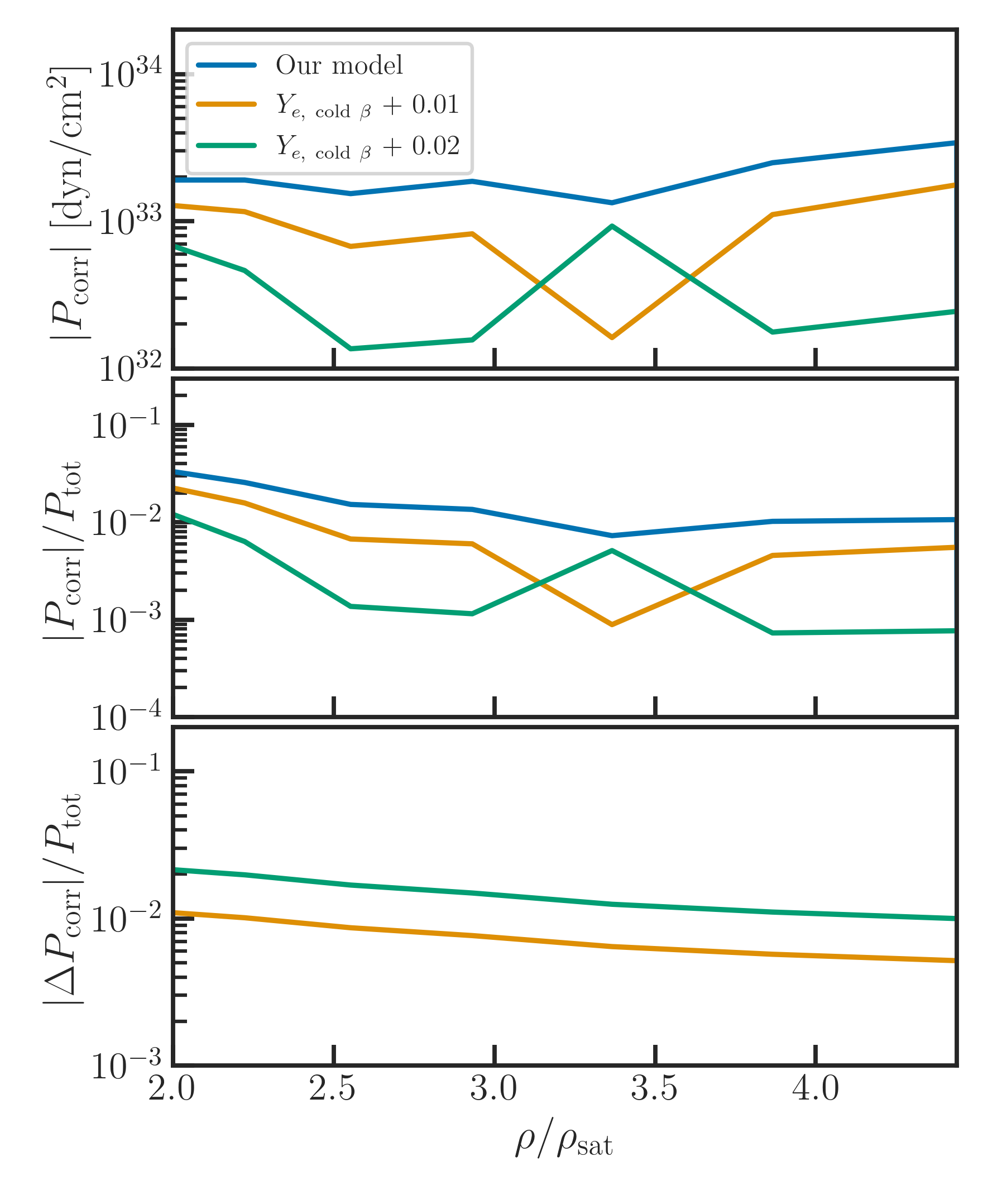} 
\caption{\label{fig:Pcorr_stiff} Same as 
Fig.~\ref{fig:Pcorr_soft}, but for the temperature
profile extracted from the late-time 
\Rbig bounce-collapse simulation. For this stiffer
case, TMA is used as the 3D EoS table for
calculating the full temperature- and composition-
dependent pressures.} 
\end{figure}

We show the results in 
Figs.~\ref{fig:Pcorr_soft} and \ref{fig:Pcorr_stiff}.
In each figure, the top rows show the
pressure corrections for the three cases 
defined in eqs.~(\ref{eq:dP_our}-\ref{eq:dP_adv}). 
The middle row shows the same pressure corrections 
relative to the total pressure. 
The bottom row shows the difference between the 
thermal-only correction assumed with our model
and the corrections 
that involve a change also to the composition, 
$ \Delta P_{\rm corr} = P_{\rm corr; ~ advection} 
    - P_{\rm corr;~ our~model}$.
We plot this difference relative to the total
pressure for the non-equilibrium composition.The reason for this normalization is that $ \Delta P_{\rm corr} = \Delta P_{\rm tot}$, where $\Delta P_{\rm tot}$ represents the change in the total pressure due to the change in composition. Therefore, the bottom panels show the fractional change in the total pressure when we account for compositional changes, 
instead of only thermal changes.

We find that, for both EoSs and both increases to $Y_e$
considered, the fractional change to the total
pressure due to composition effects is a $\lesssim2\%$
correction, compared to our model which accounts
for thermal effects alone. In other words,
for these assumptions, our thermal-only model captures $\gtrsim98\%$
of the total change to the pressure away from the
conditions of cold, $\beta$-equilibrium.

For reference, we find that changing the $M^*$ 
thermal treatment in our works leads to comparable 
(few percent) differences in $P_{\rm th}/P_{\rm cold}$ at some densities
(see Figs.~\ref{fig:medianPth_caps_R10} and \ref{fig:medianPth_caps_R14}), 
and yet such differences do not affect the threshold mass. 
Thus, as a rough approximation, we do not expect changes
to the composition to significantly affect the threshold mass.
At the least, we conclude that the thermal-only analysis
of this work is a reasonable first step towards quantifying
second-order EoS effects (going beyond the cold EoS) 
on the threshold mass for prompt collapse.

We caution that this estimate makes several assumptions,
in particular about the final $Y_e$ profiles which may not
span all possibilities.
A full analysis, including both thermal and out-of-equilibrium effects,
will be an important next step towards a final answer
on how the EoS affects the threshold for prompt collapse.

\end{document}